\documentclass[twocolumn]{aastex61}

\hypersetup{linkcolor=blue,citecolor=blue,filecolor=cyan,urlcolor=magenta}

\usepackage{natbib}
\usepackage{amsmath}

\newcommand\aastex{AAS\TeX}

\graphicspath{{./imgs/}}
\received{9. Oct. 2018}
\revised{22. Nov. 2018}
\accepted{8. Dez. 2018}

\shorttitle{\aastex\ The impact of stripped Nuclei on the SMBH density}
\shortauthors{Voggel et al.}

\begin{document}
\title{The impact of stripped Nuclei on the Super-Massive Black Hole number density in the local Universe}

\correspondingauthor{Karina Voggel}
\email{kvoggel@astro.utah.edu}

\author[0000-0001-6215-0950]{Karina T. Voggel}
\affil{University of Utah, James Fletcher Building, 115 1400 E, Salt Lake City, UT 84112, USA}

\author{Anil C. Seth}
\affil{University of Utah, James Fletcher Building, 115 1400 E, Salt Lake City, UT 84112, USA}
\author{Holger Baumgardt}
\affiliation{School of Mathematics and Physics, University of Queensland, St Lucia, QLD 4072, Australia}
\author{Steffen Mieske}
\affiliation{European Southern Observatory,Alonso de Cordova 3107, Vitacura, Santiago, Chile}
\author{Joel Pfeffer}
\affiliation{Astrophysics Research Institute, Liverpool John Moores University, 146 Brownlow Hill, Liverpool L3 5RF, UK}
\author{Alexander Rasskazov}
\affiliation{School of Physics and Astronomy and Center for Computational Relativity and Gravitation, Rochester Institute of Technology, Rochester, New York 14623, USA}




\begin{abstract}
The recent discovery of super-massive black holes (SMBHs) in high-mass ultra-compact dwarf galaxies (UCDs) suggests that at least some UCDs are the nuclear star clusters of stripped galaxies. In this paper we present a new method to estimate how many UCDs host an SMBH and thus are stripped galaxy nuclei. 
We revisit the dynamical mass measurements that suggest many UCDs have more mass than expected from stellar population estimates, which observations have shown is due to the presence of an SMBH. We revise the stellar population mass estimates using a new empirical relation between the mass-to-light ratio (M/L) and metallicity to predict which UCDs most likely host an SMBH. We calculate the fraction of UCDs that host SMBHs across their entire luminosity range for the first time.  
We then apply the SMBH occupation fraction to the observed luminosity function of UCDs and estimate that in the Fornax and Virgo cluster alone there should be $69^{+32}_{-25}$ stripped nuclei with SMBHs. This analysis shows that stripped nuclei are almost as common in clusters as present-day galaxy nuclei. We estimate the SMBH number density caused by stripped nuclei to be $2-8\times10^{-3}Mpc^{-3}$, which represents a significant fraction (8-32\%) of the SMBH density in the local Universe.
These SMBHs hidden in stripped nuclei increase expected event rates for tidal disruption events and SMBH-SMBH and SMBH-BH mergers. The existence of numerous stripped nuclei with SMBHs are a direct consequence of hierarchical galaxy formation, but until now their impact on the SMBH density had not been quantified.

   
\end{abstract}




\section{Introduction} 
\label{sec:intro}

In the hierarchical galaxy formation framework, galaxies are commonly accreted onto larger structures and their stellar content is stripped and distributed in the halo \citep[e.g.][]{Steinmetz2002, DeLucia2007}.
Several studies have shown that its possible to strip a galaxy in a larger cluster until only the central nuclear star cluster remains (\citealt{Bekki2001, Bekki2003, Drinkwater2003, Pfeffer2013, Pfeffer2014, Pfeffer2016}). A candidate for those remnant stripped nuclei are ultra-compact dwarf galaxies (UCDs) \citep{Minniti1998, Hilker1999, Drinkwater2000}. There is no physical definition of what constitutes a UCD but commonly everything brighter and more massive than $\omega$\,Cen with $2\times10^{6}M_{\odot}$ is categorised as an UCD (e.g. \citealt{Mieske2008}). As an alternative to their formation from stripped galaxy nuclei, UCDs may also be formed as massive globular clusters \citep{Murray2009, Mieske2004, Mieske2012}.

It remains unclear how many UCDs are the stripped nuclei of galaxies, but there is evidence that they are a mix of both globular clusters and stripped nuclei (\citealt{Hilker2006, DaRocha2011, Brodie2011, Norris2011}). Theoretical simulations predict that stripped nuclei make up a majority of UCDs at masses above $M>10^{7}M_{\odot}$ whereas lower-mass UCDs are predominantly GCs (\citealt{Pfeffer2014, Pfeffer2016}). Because galaxy nuclei scale with their host mass \citep{Ferrarese2006}, we can use stripped nuclei to determine the mass function of galaxies that were stripped; this would enable crucial constraints on the assembly history of a given galaxy or galaxy cluster. The presence of stripped nuclei can be used as signpost of tidal disruption and merger processes. 

One robust way to determine if a UCD is a galaxy nucleus is detecting a central super-massive BH.
holes and their host mass These have been discovered in 5 massive ($M>1\times10^{7}M_{\odot}$) UCDs (\citealt{Seth2014, Ahn2017, Ahn2018, Afanasiev2018}), whereas no SMBHs were found in two lower-mass ($M<1\times10^{7}M_{\odot}$) UCDs \citep{Voggel2018}. These UCDs with non-detections could still be either a stripped galaxy nuclei which didn't host a massive SMBH or a massive globular cluster.  
 It was predicted that tidal stripping of galaxies would generate a population of SMBHs that are over massive compared to their host galaxies mass. \citep{Volonteri2008,Volonteri2016, Barber2016}. Another cosmological simulation also indicates that numerous stripped SMBHs above $1\times10^{6}M_{\odot}$ must live in the halo of Milky Way sized galaxies \citep{Tremmel2018}.

The presence of an SMBH in the center of an UCD will increase the dynamical mass estimates derived from integrated light dispersion measurements. This is because the black hole raises the velocity dispersion of the stars near the center of the UCD and when a light-traces-mass model is assumed, the dynamical mass can be significantly over-estimated. 
Such an elevated dynamical mass has been observed in UCDs where on average the mass-to-light ratios are elevated compared to simple stellar population models (SSP) \citep{Mieske2008, Mieske2013, Strader2013}. Especially at high masses above $1\times10^{7}M_{\odot}$ most UCDs have elevated dynamical M/Ls. However, when we account for the dynamical effect of the SMBHs in the UCDs that host one, their corrected stellar $M/L$ ratios are all {\em lower} than expected from standard SSP models \citep{Seth2014, Ahn2017, Ahn2018}. 

All of these massive UCDs with SMBHs are particularly metal-rich with approximately solar metallicity, where the predicted mass-to-light ratio of SSP models \citep{BC2003, Maraston2005} increases sharply with increasing metallicity. Mass-to-light ratios that are systematically lower than SSP model prediction are also observed in lower mass globular clusters (GCs) \citep{Krui2009, Strader2011, Mieske2013, Kimmig2015}. Some works have found GC $M/L$ ratios have to be independent of metallicity \citep{Kimmig2015, Baumgardt2018}, while others even suggest decreasing $M/L$ ratios at higher metallicities \citep{Strader2011}.

In this paper we investigate whether comparing the dynamical masses of UCDs to an empirically measured M/L-metallicity relation instead of a theoretical relation is a better predictor of the presence of an SMBH in a UCD. 
Having a simple proxy for BHs in UCDs, such as the dynamical mass inferred from the integrated dispersion, would make it possible to constrain how many stripped nuclei exist in a galaxy cluster without the need for time-consuming adaptive optics observations that are not feasible for many UCDs. We use this new empirical $M/L$ relation to predict the presence of SMBHs in individual UCDs as well as a first estimate of their overall occupation fraction and their luminosity function. As a last step, we use the predicted amount of UCDs with SMBHs to estimate the increase in the number density of SMBHs in the local Universe, and quantify the effect this could have on tidal disruption and SMBH merger event rates.

\section{Empirical ML-metallicity relation} 
\label{sec:feh_ML}
Standard stellar population models predict that old stellar populations have a mass-to-light ratio that increases with metallicity. In Figure \ref{fig:feh_ML} the dashed black line shows the expected $M/L_{\rm V}$ ratio for a 13\,Gyr old stellar population, the same that was used in \cite{Mieske2013, Dabringhausen2012}, which is a mean between the predictions from \citealt{Maraston2005} and \citealt{BC2003} for a Kroupa IMF \citep{Kroupa2002}. For a stellar population of solar metallicity, it predicts $M/L_{\rm V}\sim 4.5$.

We plot GCs with dynamical $M/L_{\rm V}$ measurements from \citet{Baumgardt2018} and \citet{Strader2011} in Figure \ref{fig:feh_ML} in blue and orange respectively. The size of the circles represents the error bar on the $M/L_{\rm V}$ values. The internal dynamical evolution of GCs will cause mass segregation and the preferential ejection of low-mass stars which will decrease their $M/L$ ratio because these stars contribute little in V-band luminosity (e.g. \citealt{Vesperini1997, Baumgardt2003, Krui2009}). Thus we excluded all GCs from \citet{Strader2011} with two body relaxation times shorter than $log(t_{\rm rel})<9.3$. 

We also include data from \citet{Baumgardt2018} but we limit the sample to bright GCs with $M_{\rm V}<-7.5$ and those with a mass function slope of $\alpha<-0.8$. This eliminates clusters that have evolved strongly from the initial Kroupa mass funtion \citep{Sollima2017}; the remaining sample of dynamically unevolved clusters provides a good match to the high mass, long relaxation time UCDs. \citet{Sollima2017} found a strong correlation between mass function slope and relaxation time of a cluster. A cut-off in the mass function slope therefore ensures that we only include clusters with long relaxation times.

We then include the stellar $M/L_{\rm V}$ for 4 UCDs in green for which we have dynamical measurements of the SMBH mass {\citep{Seth2014, Ahn2017, Ahn2018}. For these UCDs the dynamical effect of the BH has been accounted for by resolving the SMBH sphere of influence, and thus the $M/L$ measurements are just for the stars.  The four UCDs all have spectroscopic metallicities close to solar \citep{Chili2008b, Firth2009, Strader2013, Sandoval2015}.

It can be seen that the theoretical trend to higher $M/L_{\rm V}$ at high metallicities is not observed in real globular clusters. Instead GCs appear to be at an almost constantant $M/L_{\rm V}$ scattered around 2. This effect has already been noted before in \cite{Strader2011} and \citet{Kimmig2015}. 
Above metallicities of [Fe/H]$>-0.5$ every single data point lies systematically below the theoretical prediction, including the metal rich UCDs.

One of the explanations for the much lower $M/L$ values of massive UCDs is that they contain a combination of an old and a young population of stars as observed in some nuclear star clusters \citep{Seth2006, Walcher2006}. Using the \citet{BC2003} models, at solar metallicity a $M/L_{V}\sim2.2$ is reached for an age of 5-6\,Gyr, younger than stellar population model fits to most UCDs \citep{Strader2013,Janz2016,Villaume2017}. A star formation history that extends over several Gyr has been found in one UCD around a field early-type galaxy NGC\,4546 \citep{Norris2015}; UCDs like this one should have significantly lower stellar M/L ratios, but represent only a small fraction of the known UCD population \citep[e.g.][]{Janz2016}.  
An initial mass function that becomes more top-heavy at higher metallicity is another possible solution to the observed difference between theoretical models and the observed M/L values \citep{Dabringhausen2009, Haghi2017}. However the very young ages (1-2\,Gyr) assumed in \citet{Haghi2017} for high-metallicity GCs seems at odds with observations that indicate that almost all metal-rich GCs in M31 are older than 5\,Gyr \citep[e.g.][]{Caldwell2011}.
Dynamical evolution also cannot account for this discrepancy between observed and theoretical M/L ratios; these UCDs are so massive that they have typical dynamical evolution timescales of $t>100$\,Gyr and thus can be considered dynamically unevolved.		

This mismatch between the observed M/Ls and theoretical predictions at high metallicity is crucial for determining inflated UCD masses. When theoretical $M/L$ predictions for UCDs are too high, UCDs which host an SMBH will not display an elevated dynamical mass.  
Regardless of why the M/L is lower in metal-rich clusters than expected (Fig.~\ref{fig:feh_ML}), an empirical relation based on similar objects should be more sensitive to the presence of BHs than the theoretical model predictions, and using our known sample of SMBHs in UCDs we show that is indeed the case below.

To avoid the mismatch between measured and theoretical M/Ls we determine an empirical metallicity-M/L relation. For this, we use the bayesian linear regression code LINMIX ERR
(\citealt{Kelly2007}) that fits a linear relation to the data. This model includes an intrinsic gaussian scatter with the witdth $\epsilon_{0}$ in the relation to account for the fact that there is a significant scatter for $M/L_{\rm V}$ values at a given metallicity.

\begin{equation}
\begin{split}
M/L_{\rm V}=(2.15^{+0.45}_{-0.53})-(0.053^{+0.45}_{-0.37}) \rm [Fe/H] \\
\epsilon_{0}=0.51^{+0.19}_{-0.14}
\end{split}
\end{equation}

This relation is plotted as black line in Figure \ref{fig:feh_ML} with the intrinsic scatter $\epsilon_{0}=0.51^{+0.19}_{-0.14}$ plotted as dotted lines. In contrast to theoretical models the empirical [Fe/H]-$M/L_{\rm V}$ relation is nearly flat, with the best-fit slope term being consistent with a flat relation.

   \begin{figure}
   \centering
   \includegraphics[width=\hsize]{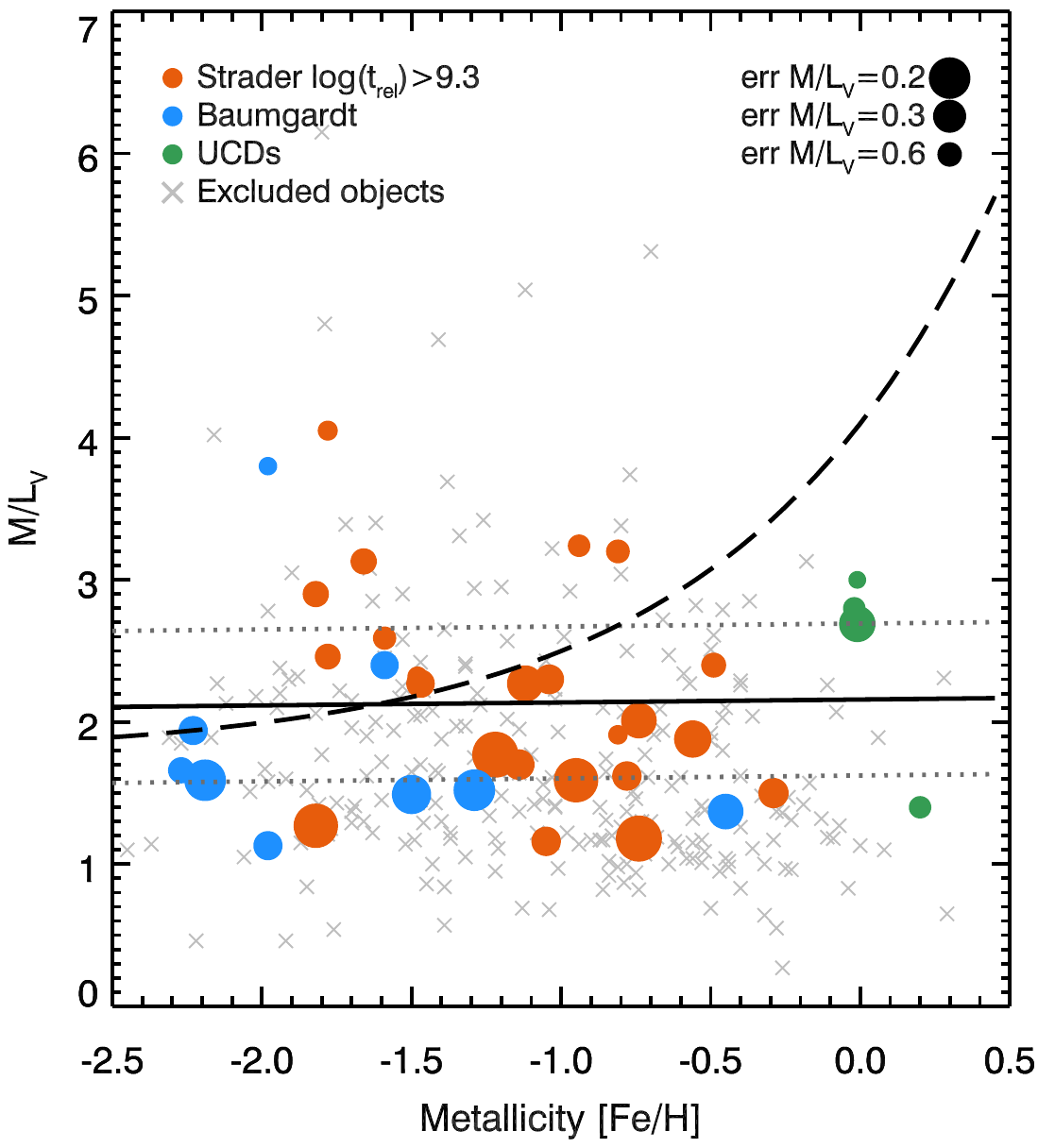}
      \caption{Metallicity [Fe/H] of GCs and UCDs plotted against their measured dynamical $M/L_{\rm V}$. We include Data from \cite{Baumgardt2017} in blue, and clusters from \cite{Strader2011} as orange points. All clusters where the two body relaxation time is longer than $log(t)>9.3$ (grey crosses) were excluded from the fit. Four UCDs \citep{Seth2014, Ahn2017, Ahn2018} for which the effect of a BH has been taken into account and their $M/L_{\rm V}$ is only from the stellar population are shown in green. The dashed black line is the mean stellar population prediction for a 13Gyr old population from \citet{Maraston2005} and \citet{BC2003}. Whereas the solid black line is out best fit linear relation and the dotted lines represent the intrinsic scatter $\epsilon_{0}=0.51$. } 
         \label{fig:feh_ML}
   \end{figure}

\section{Global $M/L$ elevation}
We reevaluate the comparison between the dynamical mass-to-light ratios of UCDs and their theoretical stellar population predictions (black dashed line Fig \ref{fig:feh_ML}), following the compilation of literature values by \citet{Mieske2013}. In the previous section we have shown that old stellar population models do not accurately reproduce the $M/L$ ratios of GCs and UCDs, and the predicted $M/L$ ratios depend critically on the model that is being used. Here we compare the dynamical M/Ls of GCs and UCDs to the empirical [Fe/H]-M/L relation we derived in the previous section.

This new comparison is based on the dynamical $M/L_{\rm V}$ values from \citet{Mieske2013} with some updates to this sample from other literature sources. Instead of using the mass on the x-axis, we use the absolute magnitude/luminosity to avoid artificial trends created by the covariance between the axes caused by errors in the integrated dispersions. We also include some changes to the sample of GCs and UCDs.  
The velocity dispersions in \citet{Taylor2010} are higher than several literature measurements and were found to be unreliable in at least one case \citep{Voggel2018}, and therefore we replace these values with the mass measurements of \citet{Rejkuba2007} for the same Cen\,A objects. We also use the same dynamically unevolved Milky Way and M31 GCs from \citet{Baumgardt2018, Strader2011} that we used in the previous section. Low-mass GCs have shorter two-body relaxation timescales that can lower their dynamical mass compared to theoretical models, thus we exclude those GCs.
 
We include updated values for UCDs where spatially resolved kinematic studies have tested the presence of a BH. These are M60-UCD1 \citep{Seth2014}, M59cO, VUCD3 \citep{Ahn2017}, M59-UCD3 \citep{Ahn2018}, Fornax-UCD3 \citep{Afanasiev2018} and UCD\,320 and UCD\,330 \citep{Voggel2018}.  We also include a new candidate for the most luminous UCD discovered in \citet{Schweizer2018}. Unlike the $M/L$ values used in the previous section, the $M/L$ values for these objects were derived from models with no BH mass included to mimic the effect of integrated light measurements made in this same way. This helps assess the robustness of using inflated dynamical M/Ls to identify BHs.

We propagate both the measured $M/L_{\rm dyn}$ error \textit{and} the intrinsic scatter of $\epsilon_{0}=0.51$ on the $M/L_{\rm pop, emp}$ when calculating the final error on the elevation quantity $\Psi=\frac{M/L_{\rm dyn}}{M/L_{\rm pop}}$. In previous work only the error on the dynamical mass was propagated \citep[e.g.][]{Mieske2013, Forbes2014} when determining how inflated a UCD is and errors on $M/L_{\rm pop}$ were not included.

   \begin{figure}
   \centering
   \includegraphics[width=\hsize]{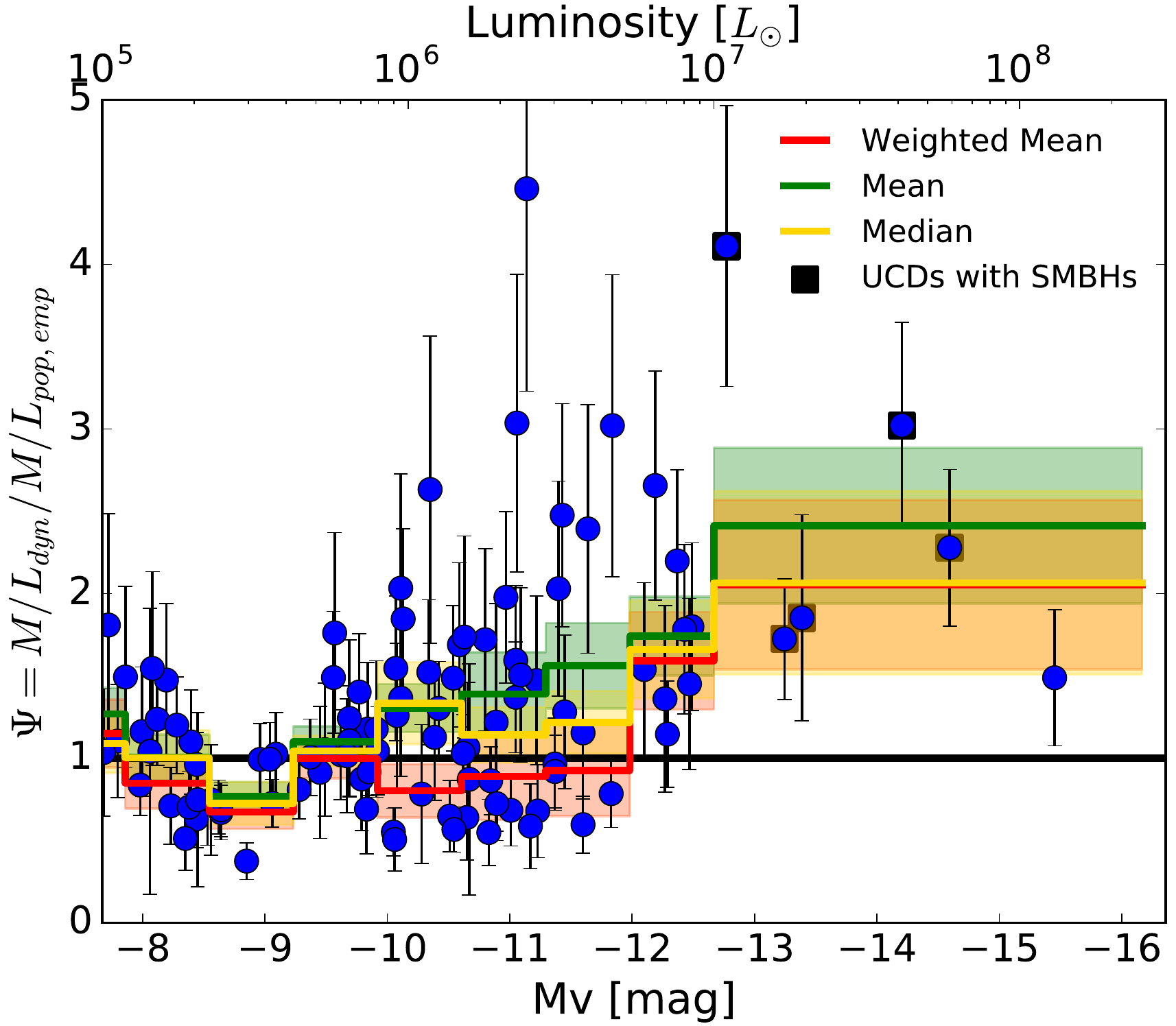}
      \caption{The luminosities of CGs and UCDs compared to their $M/L_{\rm dyn}/M/L_{\rm pop, emp}$. The mean, error-weighted mean and median average are shown in green, red and yellow respectively with the $1\,\sigma$ error as shaded area. All clusters with two-body relaxation time shorter than $log(t)<9.3$ have been excluded from this fit.} 
         \label{fig:elevation_empirical}
   \end{figure}  

The updated luminosity vs. $\frac{M/L_{\rm dyn}}{M/L_{\rm pop, emp}}$ plot is shown in Figure~\ref{fig:elevation_empirical}, where now we use the $M/L_{\rm pop, emp}$ as derived for the empirical model in the previous section. Any objects above unity (black line) are inflated compared to the empirical stellar population prediction.   
 
To determine how the average $\Psi$ changes with luminosity we bin the UCDs in 10 bins of luminosity and derive their mean (green), error-weighted mean (red) and median (yellow) elevation. The plot shows that GCs/UCDs fainter than $M_{\rm V}=-10$ are not inflated on average and all three statistical estimators agree well. However, between $-12<M_{\rm V}<-10$ the mean and median show a moderate elevation, whereas the weighted mean does not. This disagreement is driven by the fact that the M/Ls in this luminosity range are bimodal and the UCDs at very high $M/L$ values affects the mean much more as it is less outlier resistant. This bimodality has been already noted in \citet{Mieske2013}, and could result from populations of UCDs with and without BHs at these luminosities. At the bright end ($M_{\rm V}<-12$),  all UCD M/Ls and thus their averages are clearly elevated. The plot emphasizes how using standard statistical methods to determine which luminosity bins are elevated is insufficient when dealing with small samples that are not normally distributed.
The magnitude at which all UCDs have elevated M/Ls is consistent with the suggestion that true GCs become increasingly rare at brighter magnitudes, and cease to exist entirely at magnitudes above $M_{\rm V}\sim -13$ \citep{Hilker2009, Norris2011, Norris2014}. This magnitude limit also corresponds roughly to the mass of $1\times 10^{7}M_{\odot}$ where the metallicities of UCDs are exclusively high \citep{Janz2016}.

   \begin{figure}
   \centering
   \includegraphics[width=\hsize]{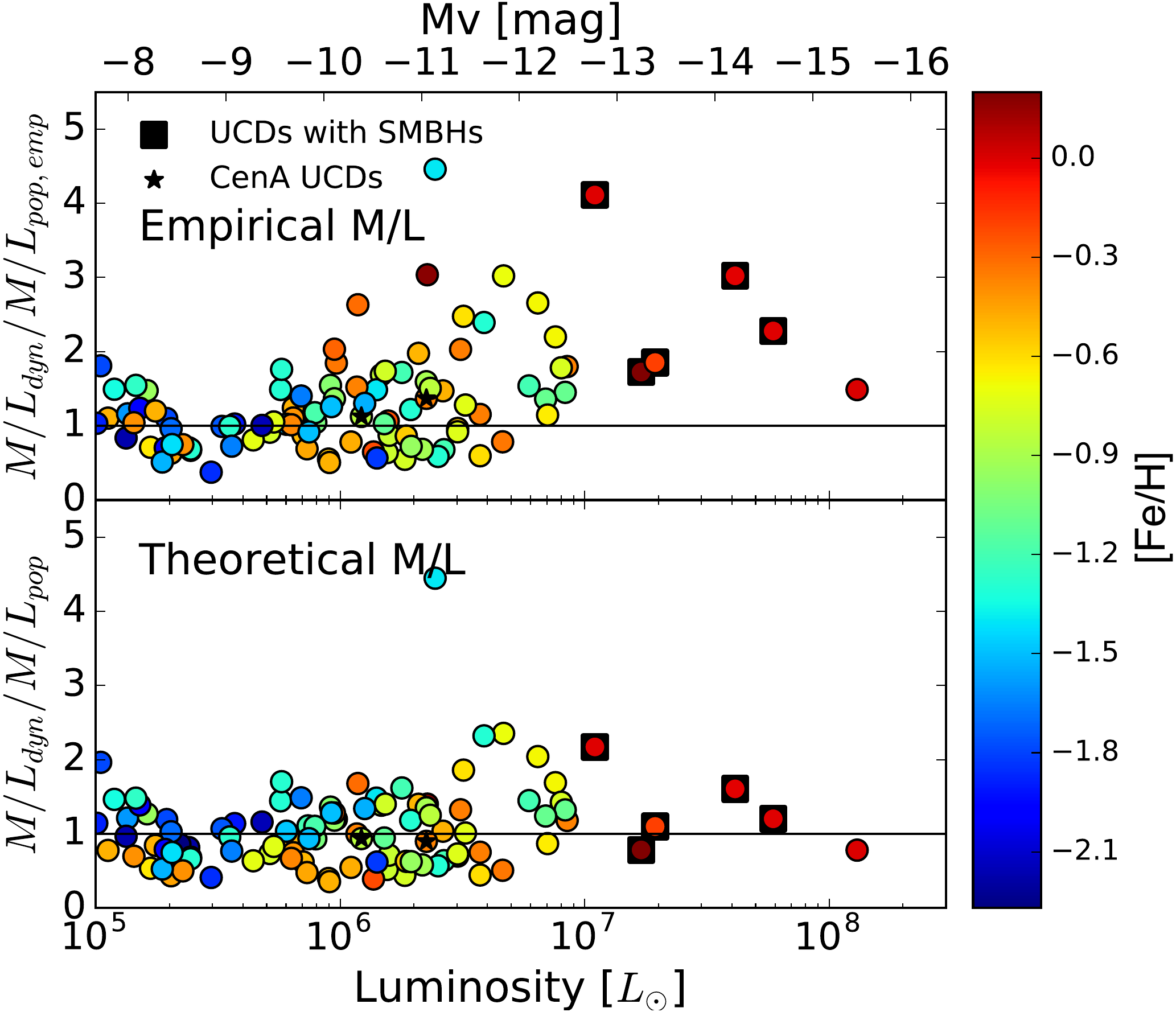}
      \caption{The $M/L_{\rm dyn}$ of UCDs and GCs compared to the new empirical theoretical $M/L_{\rm pop, emp}$ prediction in the upper panel and compared to the theoretical SSP models $M/L_{\rm pop}$ using a Kroupa IMF in the lower panel. Color coding is according to their metallicity and the black square symbols denote UCDs with confirmed BHs in their centers.} 
         \label{fig:metallicity_color_code}
   \end{figure}  

In Figure \ref{fig:metallicity_color_code} we compare the $M/L$ plot with the new empirical population prediction to the theoretical predictions. We mark UCDs with an SMBH as Diamonds. The before and after comparison shows how metal-rich objects are now more elevated than before.

With the new empirical stellar population M/L, all 5 UCDs with detected SMBHs (black diamonds, top panel Fig. \ref{fig:metallicity_color_code}) are clearly elevated. In contrast in the lower panel of Fig. \ref{fig:metallicity_color_code} most UCDs with SMBHs are not significantly inflated above $\Psi=1$, although they host SMBHs. This shows how adopting the empirical metallicity-M/L correlation instead of the theoretical relation significantly improves our ability to use inflated dynamical masses as a predictor of whether a UCD hosts an SMBH. Such a correlation between an inflated mass and SMBH presence supports the the idea that inflated dynamical masses could be used to find UCDs that host an SMBH.

\section{Individual UCDs with inflated mass-to-light ratios}

We can also use the inflation of the dynamical M/Ls of individual UCDs to predict which are the most likely to host an SMBH in their centers. To do this, we compare the $M/L$ ratio of individual UCDs to the one predicted by the new empirical [Fe/H]-M/L relation. In Fig. \ref{fig:individual_elevation} we show the significance of this elevation, which is defined as the amount the UCD is elevated above the $\Psi=1$ line in Fig.~\ref{fig:elevation_empirical}, divided by the $M/L$ error. UCDs with a measured BH in their centers are marked with orange squares, all other UCDs are shown as blue circles. 

   \begin{figure}
   \centering
   \includegraphics[width=\hsize]{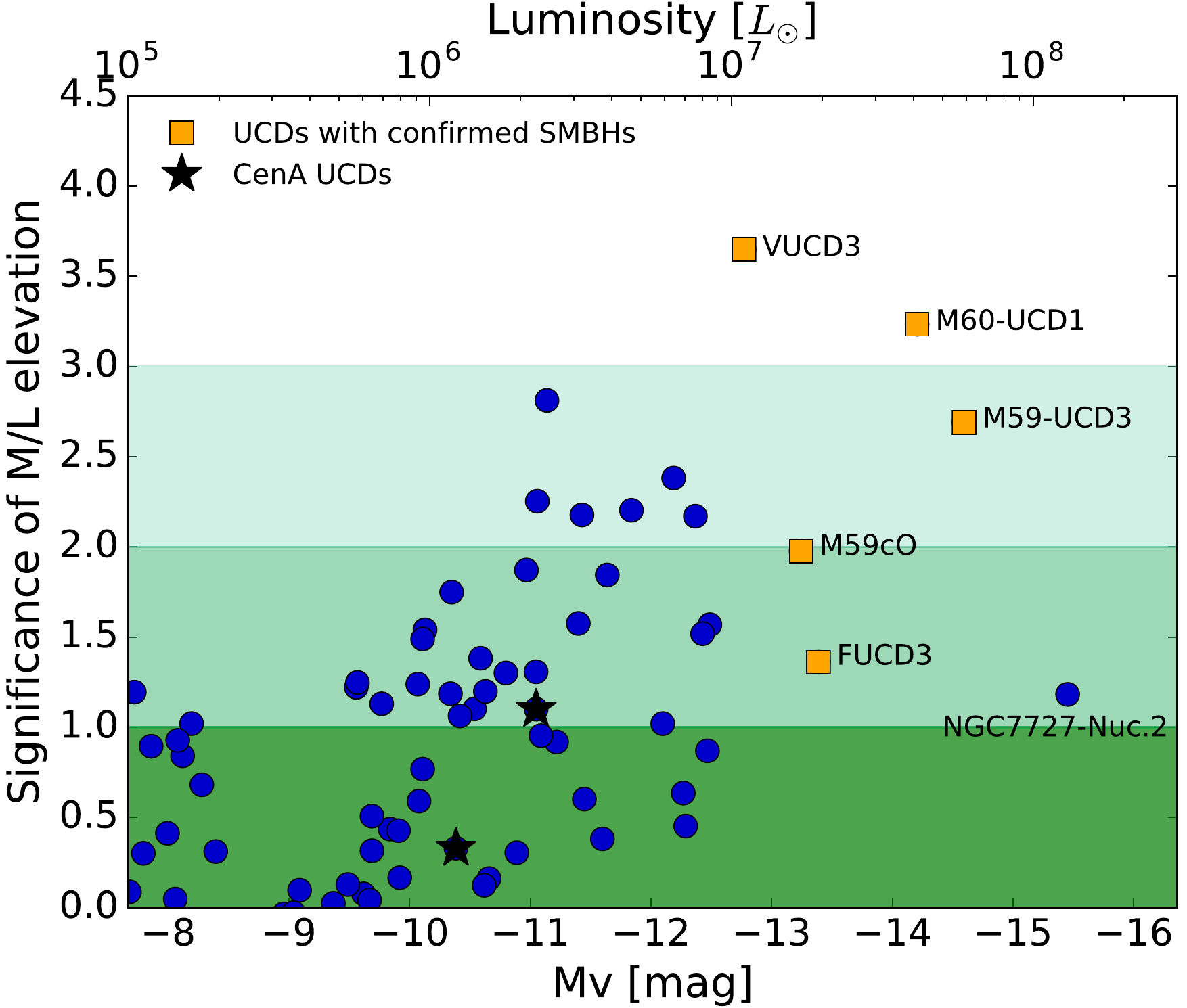}
      \caption{The significance with which individual UCDs are elevated compared to the empirical $M/L$ relation. UCDs with known BHs are marked as orange squares. UCDs with no BH measurements are shown as blue circles. The green shaded regions indicate the 1, 2 and 3\,$\sigma$ significance levels.} 
         \label{fig:individual_elevation}
   \end{figure}  

The plot shows that four UCDs with known BHs are significantly elevated at or above 2\,$\sigma$ of confidence in this plot and only FUCD3 falls slightly below. That UCDs with BHs are significantly elevated in $M/L$ using the new empirical model shows that the excess mass of individual UCDs can be used to determine whether they host SMBHs.

A list of all UCDs that are inflated above 1\,$\sigma$ and ranked after their significance is provided in Table \ref{tab:inflatedUCD}. 
All five UCDs with known BH measurements (M60-UCD1, VUCD3, M59cO, M59-UCD3 and UCD3) are elevated above 1\,$\sigma$ significance with this method and show an excess in dynamical mass. Thus how elevated an individual UCD is can be used as a way to determine those objects that are the most likely to host SMBHs and thus are the best candidates for follow-up AO spectroscopy to confirm a BH in their centers. Nevertheless with current instrumentation many UCDs at Virgo or Fornax distances are too faint for adaptive objects.  To also study the presence of SMBHs in fainter UCDs one solution is to target closer UCDs, e.g. in CenA where the AO can reach UCDs of a fainter absolute magnitude. Using this method we were able to select three UCDs in CenA with a high probability to host a massive BH for follow-up observations with SINFONI that are ongoing. We note that an additional hurdle for follow-up with IFU adaptive optics is that some UCDs have expected velocity dispersions that are below the resolution limit of typical IFU instruments such as NIFS, SINFONI and MUSE.

\begin{deluxetable}{lccccc}
\tablecaption{List of UCDs with the most significantly inflated $\Psi=M/L_{\rm dyn}/M/L_{\rm pop, emp}$ and thus the most likely targets for searching for the presence of SMBHs. They are listed in descending order of significance (col 4) and list their current $\Psi$ (col 5)and the old one using the theoretical stellar models. UCDs in bold-font are those for which an SMBH has been found. \label{tab:inflatedUCD}}
\tablehead{\colhead{Name} & \colhead{$L_{\rm V}$} & \colhead{[Fe/H]} & \colhead{Sign.} & \colhead{$\Psi_{\rm emp.}$} & \colhead{$\Psi_{\rm old}$} \\
\colhead{} & \colhead{$1\cdot10^{6}M_{\odot}$} & \colhead{[dex]} & \colhead{} & \colhead{} & \colhead{} } 
\startdata
         \textbf{VUCD3}  &  11.0  &         -0.01  &  3.65 &  4.11 $\pm$ 0.85  &  2.17 \\
      \textbf{M60-UCD1}  &  41.2  &          -0.02  &     3.23 &  3.02  $\pm$ 0.63 &  1.61 \\
          S999  &  2.4 &           -1.4  &  2.81 &  4.46 $\pm$  1.23 &  4.45 \\
      \textbf{M59-UCD3}  &  59.0  &          -0.01  &  2.69 &  2.28 $\pm$ 0.48 &  1.20 \\
          UCD1  &  6.4  &          -0.67  &   2.38  &  2.66 $\pm$  0.70  &  2.04 \\
          S490  &  2.3  &           0.18  &  2.25 &  3.04 $\pm$ 0.90 &  1.40 \\
          S417  &  4.7 &           -0.70  &  2.20  &  3.02  $\pm$ 0.92 &  2.35 \\
            F9  &   3.2 &          -0.62  &   2.18  &  2.48 $\pm$ 0.68  &  1.86 \\
           F24  &  7.6  &          -0.67  &  2.17 &  2.20 $\pm$ 0.55  &  1.69 \\
         \textbf{M59cO}  &   17.0  &        0.2  &  1.98 &  1.72  $\pm$ 0.37 & 0.78 \\
          S314  &  2.1  &           -0.5  &  1.87 &  1.98  $\pm$ 0.52  &  1.40 \\
          S928  &  3.9 &           -1.3  &  1.84  &   2.39  $\pm$ 0.76  &  2.32 \\
    HGHH92-C11  &  1.2  &          -0.31  &  1.75  &  2.63 $\pm$ 0.93  &  1.68 \\ 
            F8  &  3.1  &          -0.35  &  1.58 &  2.03 $\pm$ 0.65   &  1.32 \\       
	 VUCD5  &  8.5  &          -0.36  &  1.57  &  1.80 $\pm$ 0.51  &  1.18 \\
          0041  & 1.0 &          -0.34  &  1.54  &  1.85 $\pm$ 0.55   &  1.20 \\
         VUCD1  &  8.0  &          -0.76  &  1.52  &  1.78 $\pm$ 0.52 &  1.43 \\
    HGHH92-C29  & 0.9  &          -0.29  &  1.49 &   2.03 $\pm$ 0.69  &  1.28 \\
          0265  &  1.5  &          -0.82  &  1.38  &  1.69 $\pm$ 0.50  &  1.39 \\
        \textbf{FUCD3}  &  19.4  &          -0.19  &  1.36  &  1.85 $\pm$ 0.63  &  1.10 \\
          0365  &  2.2  &           -0.90  &  1.31  &  1.59 $\pm$ 0.45 &  1.35 \\
     HGHH92-C1  &  1.8  &           -1.20  &  1.30 &  1.72 $\pm$ 0.55 &  1.62 \\
    HGHH92-C44  & 0.6  &          -1.29  &  1.25 &  1.76 $\pm$  0.61 &  1.70\\
          0326  & 0.9  &           -1.00  &  1.24  &  1.55 $\pm$  0.44  &  1.36 \\
          0227  & 0.6 &           -1.30  &  1.22  &  1.49 $\pm$ 0.40 &  1.44 \\
    HGHH92-C17  &  1.5  &          -0.78  &  1.20 &  1.74 $\pm$ 0.61  &   1.40 \\
     B237-G299  & 0.1  &          -1.78  &  1.19 &  1.81 $\pm$ 0.68  &  1.96 \\
          0077  &  1.2  &          -0.36  &  1.19  &   1.52 $\pm$ 0.44  & 1.0 \\
NGC7727-Nucleus2  &     130.0  &    0.00  &   1.18 &  1.49 $\pm$ 0.41 & 0.78 \\
     B012-G064  & 0.7  &          -1.66  &  1.13 &  1.40 $\pm$ 0.36  &  1.49 \\
          Pal5  & 0.01  &          -1.41  &  1.11  &   2.11 $\pm$  1.01  &  2.11  \\
      VHH81-C5  &  1.4  &           -1.4  &  1.10 &  1.49 $\pm$ 0.44  &  1.48  \\
        UCD330  &  2.3  &          -0.36  &  1.10  &  1.37 $\pm$ 0.34  & 0.90  \\
       NGC5897  & 0.1  &           -1.9  &    1.09 &  1.36 $\pm$ 0.33  &  1.51  \\
       NGC5139  &  1.3  &          -1.53  &  1.06  &  1.30 $\pm$ 0.28  &  1.34  \\
       Ter8  & 0.01  &          -2.16  &  1.03  &  1.78 $\pm$ 0.76  &  2.06  \\ 
  B134-G190  & 0.2  &          -0.94  &  1.02  &  1.47 $\pm$ 0.47   &  1.27  \\
          UCD5  &  5.9 &           -1.2  &  1.02 &  1.54 $\pm$ 0.53 &  1.45  \\
\enddata
\end{deluxetable}

Although the capabilities to predict exact expected BH masses from the simple M/L-metallicity ratio are limited, we still test how well the empirical $M/L$ relation can predict BHs in those UCDs where we have detailed resolved measurements. For this comparison we retrieve the Multi Gaussian Expansion mass models of all previously published BH measurements \citep{Seth2014, Ahn2017, Ahn2018, Afanasiev2018, Voggel2018}. We then use the mass models as an ingredient in Jeans Dynamical Models (JAM) to predict their integrated velocity dispersions within their measurement aperture. We then fix the $M/L$ to the empirically predicted value for each UCD. To determine the upper and lower limit of predicted BH mass, we use the intrinsic scatter of $\epsilon=\pm 0.51$ in the metallicity-[Fe/H] relation. We then run a set of JAM models with those three fixed M/Ls and a grid of increasing BH masses. The lowest $\chi^{2}$ value for each BH mass grid is picked as predicted BH mass and its upper and lower limits. This is a way of simulating the effect of a BH on an integrated dispersion over a certain aperture. The results are plotted in Fig. \ref{fig:BH_prediction}. For the massive UCDs 3 out of 5 predictions are with the 1-$\sigma$ uncertainty range and the other two are within the 2-$\sigma$ range. For the two lower-mass UCDs with published upper-limits on their BHs \citep{Voggel2018}, the upper limits from the integrated dispersions are higher than the resolved upper limits which is expected for lower BH masses, which have a smaller effect on the integrated dispersion.

   \begin{figure}
   \centering
   \includegraphics[width=\hsize]{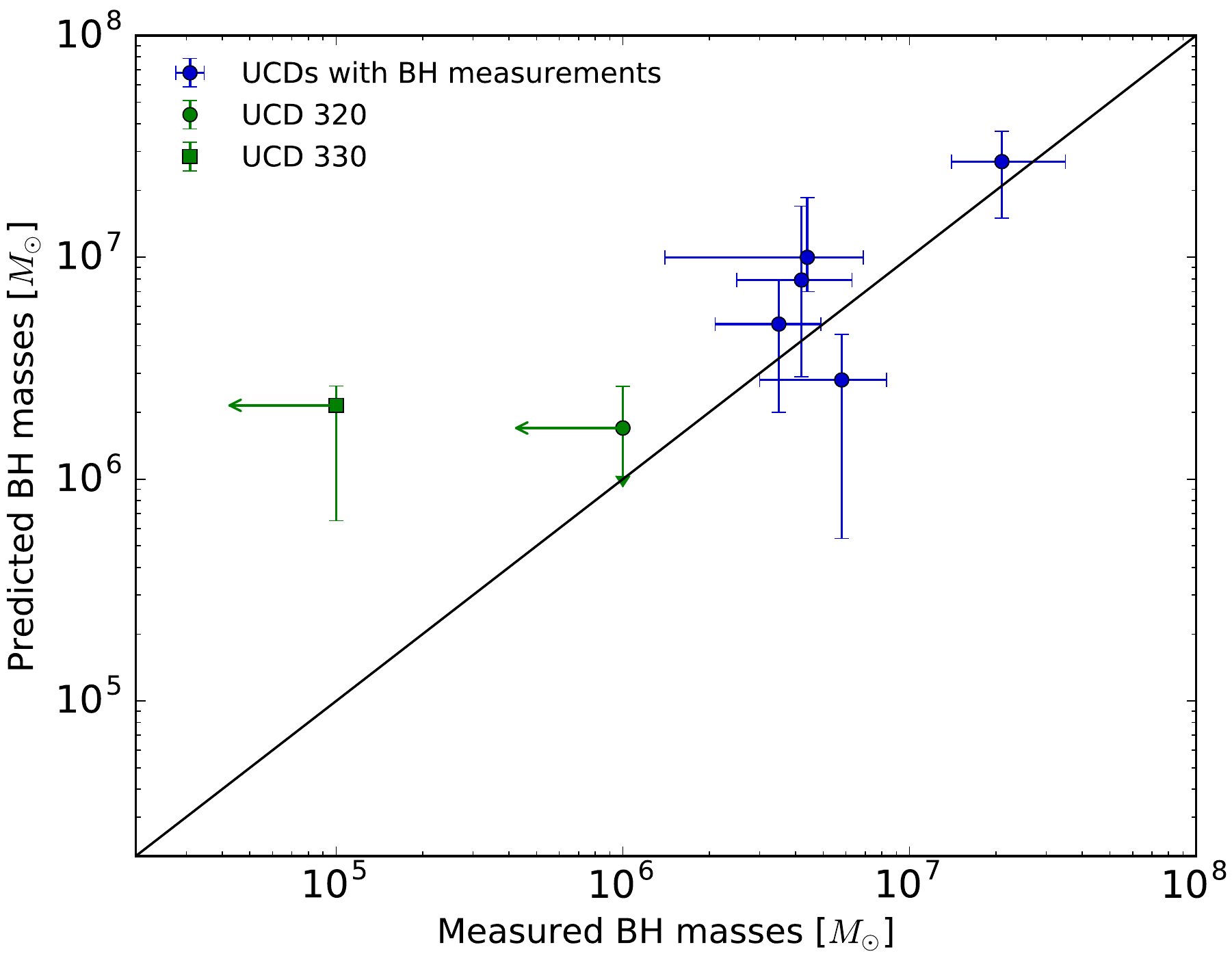}
      \caption{The measured SMBH masses in UCDs are compared to their predicted BH mass when using the new empirical $M/L$ relation. UCDs with BH measurements are shown in blue whereas the upper limits for UCD 330 and UCD320 are shown in green. The black line is the 1:1 line where the prediction is equal to the measurement.} 
         \label{fig:BH_prediction}
   \end{figure}

\section{The occupation fraction of SMBHs in UCDs}

We now turn to the overall population of SMBHs in UCDs.  SMBHs have been found in 5 massive UCDs above $10^{7}M_{\odot}$ and with $M_V < -12$ \citep{Seth2014, Ahn2017, Ahn2018, Afanasiev2018}, which all have significantly inflated $\Psi$ values in Fig \ref{fig:elevation_empirical}.
This is consistent with a picture in which an inflated $M/L$ indicated the presence of a BH. Similarly consistent with this is that we find no SMBHs in the two CenA\,UCDs that are not significantly elevated in their $M/L$ ratios (Fig,~\ref{fig:individual_elevation}).  Therefore, we derive the SMBH fraction in UCDs assuming that all UCDs with significantly inflated M/Ls do host SMBHs.

For fainter UCDs two effects likely reduce the occupation fraction of SMBHs:
\begin{enumerate}
\item We expect many GCs to mix in with stripped nuclei because at magnitudes below $M_{\rm v}=-10$ the GC luminosity function is well populated. Thus the fraction of stripped nuclei among UCDs is expected to be a function of mass, where the fraction of stripped nuclei is highest among high mass UCDs where few GCs are expected (e.g. \citealt{Pfeffer2016}).
\item Because the nuclei luminosities track those of the galaxies, the lower luminosity nuclei are expected to be from lower mass galaxies where the BH demographics are less well known, and may make up a smaller fraction of the nuclear mass \citep[e.g.][]{Graham2009, Antonini2015, Nguyen2018}. The detectability of the SMBHs is predicated on them making up a large enough fraction of the mass to be dynamically detectable from integrated spectra (which we have found to be $\gtrsim 3\%$; \citealt{Ahn2018}).
\end{enumerate}

We focus here on constraining the SMBH fraction in UCDs, which can be thought of as a lower limit on the fraction of UCDs which are stripped nuclei.

\subsection{Gaussian Mixture Model}
In this section the goal is to figure out the probability of a given UCD to have an inflated or normal $M/L$ ratio. To determine this probability, we use a Gaussian Mixture code from the Scikit-learn package \citep{scikit-learn}. These models assume that the given data is a mixture of a given number of Gaussians distributions using an expectation-maximization algorithm. 

In our implementation of the code we use the distribution of $M/L_{\rm dyn}/M/L_{\rm pop, emp}$ values (left panel Fig. \ref{fig:gaussian_mixture}) where all objects with relaxation times shorter than 2.5\,Gyr have been excluded. We determine the best fit mixture model assuming that the datapoints are drawn from Gaussian distributions. The code determines the optimal number of Gaussian distributions and their centers without the need to provide starting values. We run these Gaussian mixture models ranging from 1 to 5 Gaussian components. Evaluating the Bayesian and Akaike information criteria shows that a model with 3 Gaussians is the one that minimizes the two criteria. These two information criteria are a way of estimating the quality of a statistical model while they penalize adding new degrees of freedom. That both these criteria prefer the 3 component model while including a penalty for adding more components, suggests that three Gaussians better represent the M/L distribution of UCDs than 1 or 2 Gaussian. The probability of each $M/L$ value to belong to either of the Gaussian components is visualised in the middle panel of Figure \ref{fig:gaussian_mixture}. 

The clustering code picks up the blue component which is centred at an  $M/L_{\rm dyn}/M/L_{\rm pop, emp}$ of 1 which would correspond to the distribution of non-inflated GCs. A second peak at  $M/L_{\rm dyn}/M/L_{\rm pop, emp}\sim 2$ in red is what are likely the distribution of UCDs with SMBHs. 

The two components at large $M/L_{\rm dyn}/M/L_{\rm emp.}$ may just be due to a non-Gaussian distribution of SMBH masses. However, the third component could also be due to extremely elevated $M/L$ ratios if they are in the process of being tidally stripped and thus not in dynamical equilibrium (see e.g. \citealt{Forbes2014, Janz2015}).
For now, we assume that all inflated M/Ls are due to SMBHs, as the lifetimes of the very inflated M/Ls during tidal stripping are short, and thus these objects are expected to be rare relative to fully stripped nuclei in dynamical equilibrium.

   \begin{figure*}
   \centering
   \includegraphics[width=\hsize]{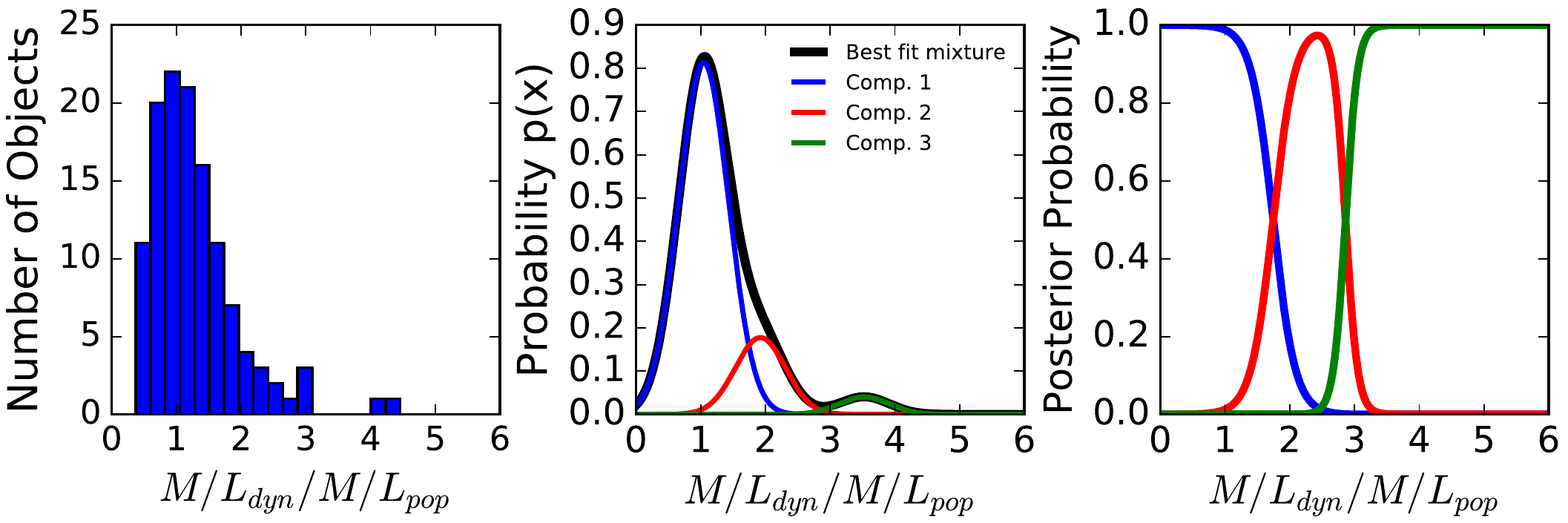}
      \caption{Left Panel: The histogram of the distribution of $M/L_{\rm dyn}/M/L_{\rm pop, emp}$ values. In this histogram all objects with relaxation times of $log(t_{\rm rel})<9.3$. Middle Panel: The probability of the best fit Gaussian Mixture model is shown in black and the three mixture components in blue, red and green respectively. Right Panel: The posterior probability that a given UCD was drawn from one of the three components as a function of $M/L_{\rm dyn}/M/L_{\rm pop, emp}$. The colour scheme is the same as in the middle panel. } 
         \label{fig:gaussian_mixture}
   \end{figure*}

Now that we have established a way to assign a probability to each individual UCD of whether it belongs to either the inflated or non-inflated $M/L$ categories, we can apply it to the overall UCD population for each luminosity bin. While for an individual UCD the $M/L$ is a noisy measurement, the statistics for the overall population of UCDs could provide a first estimate of the SMBH occupation fraction in UCDs and how it varies with luminosity.

We use the same luminosity bins as in Fig. \ref{fig:elevation_empirical} and and take the average probability of each object in that bin to belong to either of the two inflated (red and green in Fig. \ref{fig:gaussian_mixture}) components. This distribution of the average probabilities to belong to the inflated components is shown in Fig. \ref{fig:occupation_fraction} as a function of luminosity. This average probability can be interpreted as the occupation fraction of SMBHs in UCDs under the assumption that an inflation corresponds to the presence of an SMBH. 

The error on the occupation fraction (blue shaded region) is determined by doing a bootstrapping and error resampling to account for the individual errors on the $\Psi$ and the intrinsic small sample size in each bin. We do this by doing 100 runs of error resampling each $M/L$ value and then drawing randomly from the underlying $M/L$ distribution the same amount of measurements, and thus performing a replacement resampling. To each of the 100  resampled $M/L$ datasets we then apply the same method as above to determine the average occupation probability and the standard deviation between these 100 resampling is the error.

   \begin{figure}
   \centering
   \includegraphics[width=\hsize]{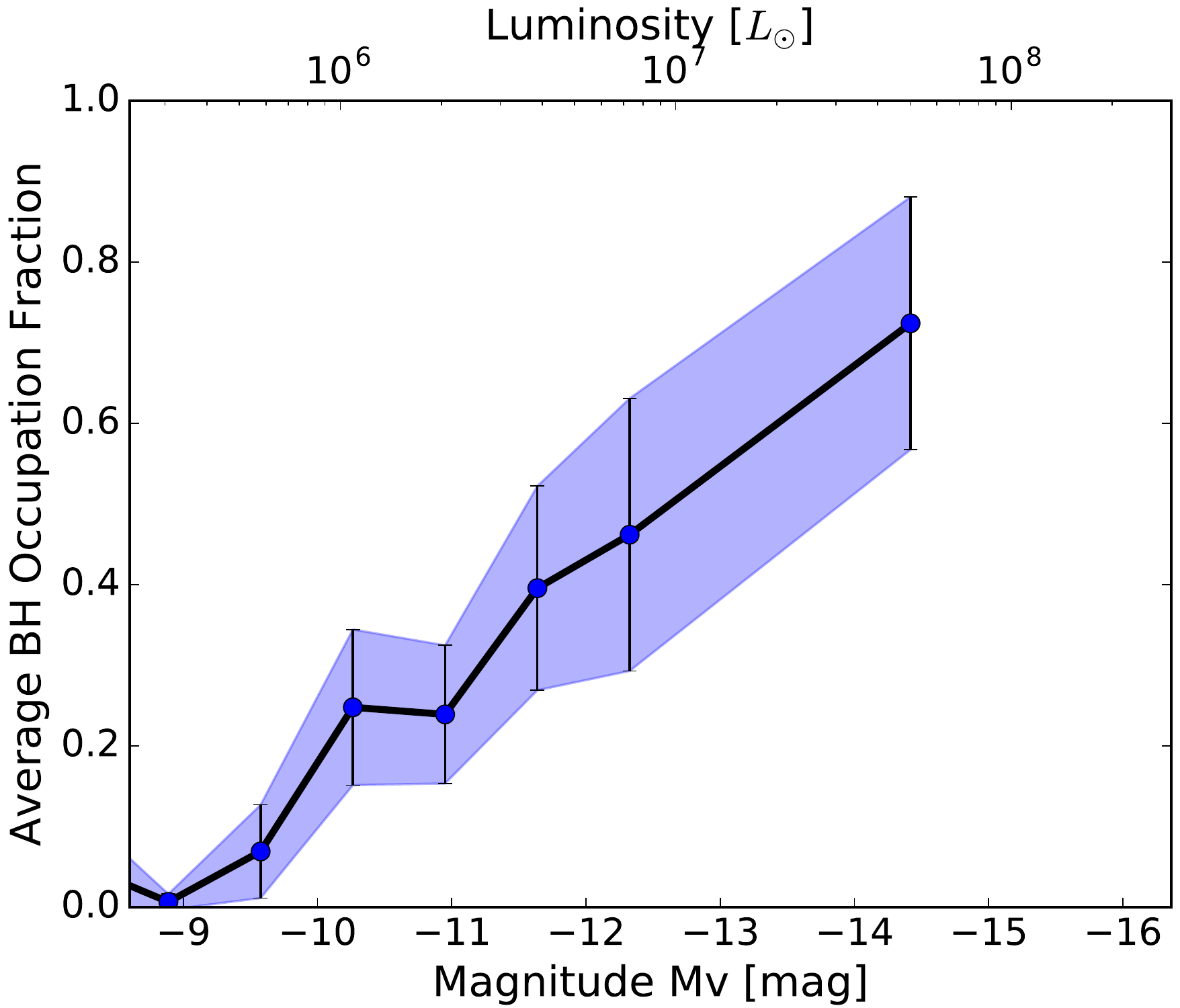}
      \caption{The average occupation probability of SMBHs in UCDs as function of their luminosity. Here it was assumed that any object belonging to the two elevated $M/L_{\rm dyn}/M/L_{\rm emp}$ categories of the Gaussian Mixture Model hosts a BH. The blue shaded region indicates the error of the SMBH occupation fraction which is a combination of doing an error resampling and a sample resampling.} 
         \label{fig:occupation_fraction}
   \end{figure} 

 As expected the occupation probability of SMBHs is small at the low-mass end where GCs are likely dominating ranging between 0-15\% at magnitudes fainter than $M_{\rm V}=-10$. In the luminosity bin between $-10>M_{\rm V}>-12$ the occupation fraction rises to values between 20-40\% and at the brightest magnitudes it is between 45-80\%.  These high occupation fractions in the highest luminosity bin are consistent with the observations that all 5 UCDs in this luminosity range where measurements have been made do in fact have SMBHs \citep{Seth2014,Ahn2017,Ahn2018,Afanasiev2018}. 
This is the first observational quantification of the occupation fraction of SMBHs in UCDs.

\section{The Number of Stripped Nuclei in Local Galaxy Clusters}
\label{sec:strippednuclei}
We now use the observed UCD population in the Virgo and Fornax clusters to predict the number of SMBHs that could be hidden in UCDs in the local Universe. We do this by multiplying the observed UCD luminosity functions with our derived estimate of the SMBH occupation fraction (Fig.\,\ref{fig:occupation_fraction}). From this we get a luminosity function of expected UCDs with SMBHs that are the former nuclei of galaxies. In a second step we can then compare these stripped nuclei to the current number of nuclear star clusters. With this number we can determine how many galaxies were already stripped onto a galaxy cluster compared with its present-day galaxy content.

   \begin{figure}
   \centering
   \includegraphics[width=0.9\hsize]{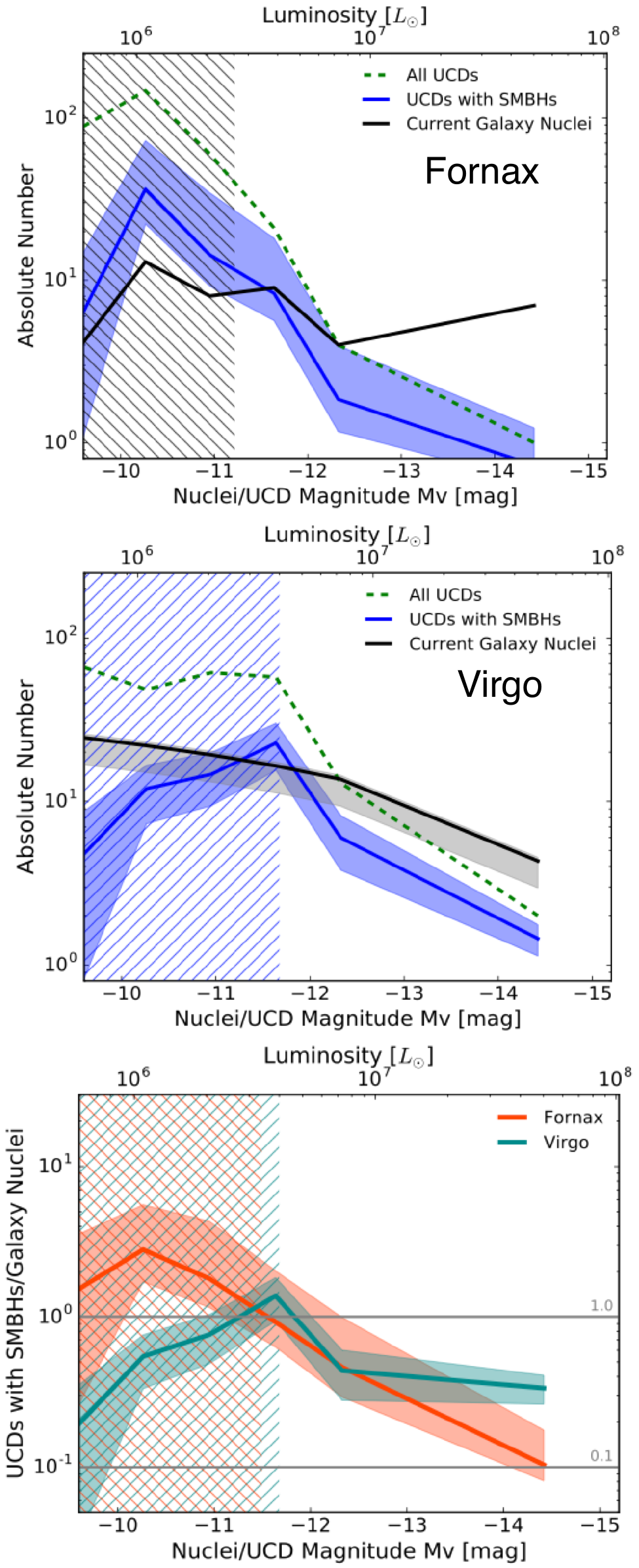}
\caption{{\em Top two panels:} The overall luminosity function of UCDs in Virgo (top) and Fornax (middle) are shown as green dashed lines, UCDs with SMBHs as blue lines, and the number of current galaxy nuclei as black lines. The uncertainties in the number of UCDs with SMBHs and galaxy nuclei are the shaded blue and grey areas. The luminosity at which the samples are incomplete is marked with a hatched black region (Fornax) and a hatched blue Region (Virgo). {\em Bottom Panel:} The ratio of the number of UCDs with SMBHs divided to the number of current galaxy nuclei. The 1.0 and 0.1 levels are marked as grey lines. The hatched areas correspond to the magnitudes at which incompleteness affects the two samples.} 
         \label{fig:nuclei_ucds}
   \end{figure}  
   
   For the Fornax Cluster we use the catalogue of \citet{Hilker2007} containing 325 spectroscopically confirmed GCs and UCDs. The catalogue has a completeness of $\sim70\%$ within 50\,kpc and drops to 30-50\% beyond 100\,kpc (see \citet{Mieske2012} for a discussion of the completeness). We use the completeness estimate to calculate the upper limit on the number of expected UCDs in Fornax. 
   
   The galaxy nuclei in Fornax are a combination of two catalogues. For nucleated galaxies brighter than $M_{g}=-16$ we use the nearly complete sample of 43 nucleated early type dwarf galaxies from \citep{Turner2012} distributed out to the virial radius of the cluster ($\sim$1.3~Mpc). They note that their detection limit for nuclei in these galaxies is $L_{\rm V}=3.3\times10^{6}L_{\odot}$ which is marked by the hatched area in the top panel of Fig.\,\ref{fig:nuclei_ucds}. At the faint end ($M_{\rm g}>-16$) of the galaxy luminosity function we use the study of \citet{Ordenes2018} that analyzed 61 nuclei in these faint galaxies \citep{Munoz2015} located within the central 350\,kpc of the Fornax cluster; however most of these objects have nuclei which are fainter than the UCDs we consider here.  For our comparison we combine both catalogues to get a dataset that is as complete as possible.  
   


In the Virgo cluster the samples of UCDs and nucleated galaxies are less complete.  For the Virgo Cluster UCDs we use the \citet{Zhang2015} Catalogue, which contains 97 spectroscopically confirmed UCDs within $\sim$600~kpc of M87, the center of the Virgo\,A subcluster. In addition to a complex selection function that results from compiling several studies, this catalogue also places a lower size limit of 10\,pc on what they define as a UCD which likely causes incompleteness at the faint end. We attempt to correct for this incompleteness. The mass-size relation of UCDs (e.g. \citealt{Norris2011}) reaches 10\,pc at a luminosity of $\sim5\times10^{6}L_{\odot}$. Below this luminosity the \citet{Zhang2015} should be missing a significant number of more compact objects. To estimate the fraction of UCDs at these luminosities that are smaller than the 10\,pc size cut in \citet{Zhang2015} we use the catalogues of bright UCDs/GCs in Centaurus\,A with size measurements from \citet{Rejkuba2007} and \citet{Taylor2015}. For luminosities between $1\times10^{6}L_{\odot}<L<5\times10^{6}L_{\odot}$ we find that 53.8\% of UCDs have effective radii that are smaller than 10\,pc. We thus correct our number of UCDs in the Zhang sample with this incompleteness for every luminosity bin below $L<5\times10^{6}L_{\odot}$.  Despite this correction, the luminosity function of UCDs flattens at the faint end -- we suspect this is due to incompleteness in the heterogeneously selected sample of spectroscopically confirmed objects around M87.

No complete catalogue of nucleated dwarf galaxies exists in Virgo as it does in Fornax. Thus we use the observed Virgo core galaxy luminosity function from \citep{Ferrarese2016} to derive the current number of nuclear star clusters; this luminosity function roughly matches the areal coverage of the UCD sample above.
Their best-fit early type galaxy luminosity function is a Schechter function with a slope of $\alpha=-1.33$. We normalize it to contain exactly the N=404 galaxies they find within their magnitude limit range to derive the number of galaxies per luminosity bin and  then multiply it with the galaxy nucleation fraction (their Fig. 7), which we approximate as a linear relation that drops from 95\% at $M_{\rm g}=-20.45$ to close to 0 for low luminosity galaxies at $M_{\rm g}=-10.0$. The error bar on their nucleation fraction is directly propagated in the number of nucleated galaxies. Now we have the number and luminosity function of nucleated galaxies in Virgo, and as a last step we use the galaxy nuclei luminosity relation $L_{\rm Nuc}=0.0032 \times L_{\rm Gal}$ from \citep{Cote2006} to derive the estimated magnitude of the nuclei themselves. 

We summarize our comparison of UCDs and current galaxy nuclei in Fig.~\ref{fig:nuclei_ucds}.  The luminosity functions of both Fornax and Virgo nucleated dwarf galaxies are shown as solid black line in the top and middle panels of~Fig. \ref{fig:nuclei_ucds}, whereas the number of total UCDs is shown as green dashed line. The black hatched region marks luminosities below $L_{\rm V}=3.3\times 10^{6}L_{\odot}$ where the observed galaxy nuclei in Fornax are potentially incomplete in \citet{Turner2012}. The blue hatched region in the Virgo panel marks the luminosities below $L_{\rm V}=4\times 10^{6}L_{\odot}$ where the UCDs in Virgo are likely affected by incompleteness due to the 10\,pc size cut.

\subsection{Comparing UCDs to present-day nuclei}
We now calculate the luminosity function of UCDs hosting SMBHs by multiplying the UCD luminosity function with the occupation fraction from Fig. \ref{fig:occupation_fraction}.  This is added as the blue line and shaded region in Fig. \ref{fig:occupation_fraction}.  	
The error on the number of UCDs with SMBHs (blue shaded region) is determined by propagating the uncertainty in the derived occupation fraction to the expected SMBH numbers.  In Fornax, the upper limit on the number of UCDs with SMBHs also includes the completeness correction based on the estimated completeness of UCDs in \citet{Mieske2012}. In both galaxy clusters, the luminosity function of UCDs with SMBHs (blue) shows a decreasing trend towards the brightest object.

At the bright end, the number of present-day nuclei outnumbers the number of UCDs by a factor of a few, however, at fainter luminosities the ratio of UCDs to galaxies appears to rise.  We compare this ratio in both clusters direction in the bottom panel  of Fig. \ref{fig:nuclei_ucds}. Taking Fornax, where there is the sample of galaxies is quite complete, UCDs appear to outnumber present-day nuclei between $M_{\rm V}$ of -10.5 and -11.5, with incompleteness possibly affecting our nuclei sample in the low luminosity bins.  

Virgo provides a consistent picture at the bright end with UCDs outnumbering present day nuclei at $M_{\rm V}=-11.5$, but then declines towards fainter magnitudes, likely due to incompleteness.

Summing up all UCDs with SMBHs above a luminosity of $L_{v}>1.5\times10^{6}L_{\odot}$, we expect $25^{+16}_{-9}$ in the Fornax cluster, and $44\pm16$ in the Virgo cluster. From this correlation we expect that there are at least $69^{+32}_{-25}$ SMBHs hidden in UCDs in the local clusters Fornax and Virgo. In comparison we currently know of 28 galaxy nuclei of the same luminosity in Fornax and 54 galaxy nuclei in Virgo. This is a ratio of SMBH UCDs vs. current nuclei of $0.89^{+0.57}_{-0.32}$ in Fornax and $0.83^{+0.22}_{-0.37}$ in Virgo respectively. While these numbers have high uncertainties they indicate that we can expect to find one SMBH hidden in a UCD for each current nucleated galaxy in our two nearest galaxy clusters. 

In addition to comparing the expected SMBHs in UCDs with the present day galaxy nuclei we also compare these numbers to the known SMBHs in the Local Universe. For this we use the compilation of 97 measured SMBH masses in the Local Universe from \citep{Saglia2016}. Of these SMBHs 46 are at distances of D$<20$\,Mpc. This is a lower limit on the amount of SMBHs in the local Universe as not every galaxy has a dynamical measurement available. Using the total amount of predicted SMBHs in Fornax and Virgo UCDs of 69 we derive a relative ration of SMBHs in UCDs vs. present day known SMBHs of $69/46=1.5$. This could be viewed as an upper limit on the UCD BH vs. existing BH fraction. However, the number of UCDs is also likely underestimated, as we have included only those in the two massive clusters within 20 Mpc, while there are known UCDs with inflated M/Ls around other galaxies (e.g. CenA and Sombrero).  We estimate the number density of SMBHs in UCDs in the next section.

\section{Discussion}
\label{sec:discussion}

\subsection{The contribution of stripped galaxy nuclei to the local number density of SMBHs}
We use the occupation fraction to estimate the total number density of UCD BHs in the local Universe. We make two estimates on the number density of UCD SMBHs:

\begin{enumerate}

\item As a first estimate of the UCDs with SMBHs in the local Universe, we take the total estimated number of Fornax and Virgo UCDs with SMBHS and divide by the volume (D$<$20\,Mpc) to which both clusters extend. The total number of predicted UCDs with SMBHs above $L_{v}>1.5\times10^{6}L_{\odot}$ in Fornax is $25^{+16}_{-9}$  and and $44\pm16$ in Virgo (blue lines in Figure \ref{fig:nuclei_ucds} and thus a total of $69^{+32}_{-25}$ UCDs with SMBHs are expected in both clusters. Using a sphere with a radius of 20\,Mpc this leads to a UCD SMBH number density of $n = 2.1^{+0.98}_{-0.75}\times10^{-3}\frac{\rm SMBHs}{\rm Mpc^{3}}$ that are hidden in UCDs. This is a lower limit on the presence of hidden SMBHs assuming that these stripped nuclei only exist in massive clusters in the Local Volume.

\item For a less conservative estimate we estimate the total number density of early type galaxies in the same volume, and assume the UCD SMBH/nuclei ratio we find here applies in all environment (not just cluster environments). For this we use the local number density measured in \citep{Blanton2005} of early type galaxies. They find an approximately constant value of $\Phi\sim3.5\times10^{-3}\rm mag^{-1}Mpc^{-3}$ in the galaxy magnitude range between $-13.5<M_{\rm r}<-20.0$. We multiply this galaxy luminosity function with the nucleation fraction from \citep{Ferrarese2016}  to determine how many nucleated galaxies are expected in each magnitude bin. 

We then use this luminosity function of nucleated galaxy and apply the galaxy-to-nuclei luminosity relation $L_{\rm Nuc}=0.0032 \times L_{\rm Gal}$ from \citep{Cote2006}. Now we have the luminosity function of the nuclei themselves and can multiply them with the derived UCD SMBH/nuclei ratio (see last panel Fig. \ref{fig:nuclei_ucds}), to derive the luminosity function of predicted stripped nuclei. When summing all predicted stripped nuclei brighter than $L_{v}>1.5\times10^{6}L_{\odot}$, we get a total number density of SMBHs in UCDs of $n=7-8\times10^{-3}\rm Mpc^{-3}$ based on our  Fornax and Virgo cluster data. 

\end{enumerate}

Combining these methods we get estimates for the number density of SMBHs in UCDs in the range of $2-8\time10^{-3}\,\rm SMBHs/Mpc^{3}$.  The total number density of SMBH above $10^{6}\,M_{\odot}$ in local galaxies was determined by \citet{Shankar2004} to be 0.017\,Mpc$^{-3}$.  Recently it has become clear that black holes down to $10^5\,M_{\odot}$ are common in lower mass galaxies \citep{Reines2013,denBrok2015,Nguyen2018,Chilingarian2018} and many of our UCDs likely have BHs $< 10^{6}\,M_{\odot}$; we therefore integrate the \citet{Shankar2004} BH mass function down to $10^{5}M_{\odot}$ to a get an SMBH number density estimate of 0.025\,Mpc$^{-3}$. The SMBHs in UCDs represent a significant increase (8-32\%) to the total SMBH number density.


\subsection{The impact of SMBHs in UCDs on TDE and SMBH merger event rates}

The presence of BHs in UCDs has a number of implications for other astrophysical phenomenon, including mergers of stellar mass BHs with SMBHs and tidal disruption events (TDEs).
Nuclear star clusters are a major source of stellar-mass black hole gravitational captures -- close encounters between two single BHs leading to gravitational wave energy loss and subsequent BH binary formation \citep{OLeary2009}. Since the event rate per galaxy is almost independent of the SMBH mass \citep{Kocsis2012,Gondan2018a} and also dynamical friction is more efficient at bringing BHs into the galactic center in small galaxies \citep{Rasskazov2018}, the total event rate is dominated by lower mass BHs and galaxies. The existence of SMBHs in UCDs increases the total density of SMBHs by 8-32\% and thus consequently also the event rate by the same fraction. Different estimates of the total GW capture event rate span the range  0.02 -- 1 $\rm yr^{-1}Gpc^{-3}$ \citep{OLeary2009,Tsang2013,Rasskazov2018}. Even if that channel doesn't produce the majority of LIGO events, most of GW captures have high eccentricities measurable by the Advanced LIGO-VIRGO-KAGRA detector network \citep{Gondan2018b} which makes it potentially possible to distinguish them from the other BH binary formation channels.

Tidal disruption events (TDE) numbers are dominated by the low mass end of the SMBH mass function. In fact, most TDEs are observed to be in relatively low luminosity galaxies \citep{LawSmith2017} and at low BH masses \citep{Wevers2017, Mockler2018}.  
Given our findings of comparable numbers of stripped and present day nuclei, we might expect a large number of TDEs to be occurring outside of present-day galaxy nuclei.  Note that at the distance of many TDEs, UCDs may not be detectable. However, recent results suggest TDEs populate galaxies with post-starburst spectra preferentially \citep{French2016} if recent star formation enhances TDE rates, then UCDs, with their old populations, may host fewer TDEs than expected based on their relative number of SMBHs.

\section{Conclusions}

This paper focuses on trying to understand the population of stripped nuclei hidden amongst UCDs.  For inferring the presence of stripped nuclei we use estimates of the dynamical M/L ratio from integrated dispersions, which can be inflated due to the presence of SMBHs in the UCD; we assume UCDs with inflated M/Ls are stripped nuclei. 

We find the following:

\begin{enumerate}
\item that UCDs and GCs appear to be best fit by a nearly constant $M/L_{\rm V}$ vs. [Fe/H] relation; this suggests that the $M/L_{\rm V}$ is significantly lower than stellar population models at the metal-rich end.
\item  all five UCDs for which a SMBH was found with high-resolution adaptive optics data correctly show an excess in dynamical mass when using the new empirical $M/L_{\rm V}$-metallicity relation.
\item we can identify additional candidates significantly enhanced dynamical M/Ls; while most are too faint to be observed with current adaptive optics instruments.
\item 
using a Gaussian Mixture Model we can predict how many UCDs in each luminosity bin are expected to host an SMBH and thus be the stripped nuclei of galaxies. We use this Mixture Model to estimate how the occupation fraction of UCDs with SMBHs vary with their luminosity. We find a small occupation fraction at low luminosities ($<$20\% at $M_{\rm V} < -10$), rising to $\sim$75\% at $M_{\rm V} = -14.5$. While stripped nuclei dominate the luminosity function at the bright end, at the lower luminosity end the stripped nuclei are mixed with ordinary GCs, which naturally causes a bimodality in the $\Psi$ distribution. 

\item comparing UCDs to present day nuclei in Virgo and Fornax, we find that stripped nuclei with SMBHs are almost as common ($\sim$90\%) than present-day galaxy nuclei and  would potentially double the density of galaxy nuclei and thus SMBHs in the local Universe.  
The ratio of UCDs to galaxy nuclei appears to decrease with increasing luminosity.
\item the number density of SMBH in UCDs is high which will significantly increase the density of SMBHs in the local Universe. Our number density estimates of UCDs with SMBHs range from $2-8\times10^{-3}Mpc^{-3}$. Stripped nuclei could increase the SMBH density - and thus tidal disruption as well as binary BH LIGO merger rates by 8-32\%. Due to their significant impact, it is important to take SMBHs in UCDs into account when determining event rates.

\end{enumerate}

We conclude that while all these numbers are first estimates it is clear that the contribution of stripped galaxy nuclei that are non-negligible. Those "hidden SMBHs" in stripped nuclei will significantly increase the amount of SMBHs in the local Universe as they approach the number of current galaxy nuclei. The existence of stripped galaxy nuclei is a direct consequence of hierarchical galaxy formation. However as these old nuclear star clusters are mostly devoid of gas they are not detectable with conventional Radio/X-ray searches and thus have been missed until now. 

If we can reliably find UCDs that are the nuclear star clusters of stripped galaxies then they have the potential to be a unique direct tracer of the past accretion history of a galaxy cluster. However, currently, the easiest way to securely identify a stripped nucleus is through identification of an SMBH (but see \citet{Norris2015} for alternative methods). Such observations require high spatial resolution adaptive-optics IFU observations that are only possible for the very brightest UCDs and those that are nearby. Until the advent of 30m class telescopes, such as the ELT, these requirements severely limit and bias the sample of UCDs for which this measurement is even possible. Therefore it is of high importance to measure the BH mass in more UCDs across the mass range for which these observations are still feasible. This will then help to establish whether there is a successful indirect proxy, such as the integrated dynamical mass used in this paper, which then can be used to trace stripped former nuclear star clusters and thus their hierarchical accretion for a given galaxy cluster on a large scale.


\acknowledgments
Work on this project by K.T. V. and A.C.S. was
supported by AST-1350389.
JP gratefully acknowledges funding from a European 
Research Council consolidator grant (ERC-CoG-646928-Multi-Pop).
A. R. received funding from the European Research Council (ERC) under the European Union's Horizon 2020 Programme for Research and Innovation ERC-2014-STG under grant agreement No. 638435 (GalNUC).
We thank the referee for their careful reading of the draft and helpful suggestions to improve the paper.

%

\vspace{5mm}
\software{This research made use of Astropy,\footnote{http://www.astropy.org} a 
community-developed core Python package for Astronomy \citep{astropy:2013, astropy:2018}. 
This research has made use of NASA's Astrophysics Data System, the Scikit-learn \citep{scikit-learn} code, 
the SciPy \citep{jones_scipy_2001} package and NumPy \citep{van2011numpy}. }





\bibliographystyle{apj}
\bibliography{bibliography_metallicity}

\begin{thebibliography}{}
\expandafter\ifx\csname natexlab\endcsname\relax\def\natexlab#1{#1}\fi

\bibitem[{jon(2001)}]{jones_scipy_2001}
 2001, {SciPy}: Open source scientific tools for Python

\bibitem[{{Afanasiev} {et~al.}(2018){Afanasiev}, {Chilingarian}, {Mieske},
  {Voggel}, {Picotti}, {Hilker}, {Seth}, {Neumayer}, {Frank}, {Romanowsky},
  {Hau}, {Baumgardt}, {Ahn}, {Strader}, {den Brok}, {McDermid}, {Spitler},
  {Brodie}, \& {Walsh}}]{Afanasiev2018}
{Afanasiev}, A.~V., {Chilingarian}, I.~V., {Mieske}, S., {et~al.} 2018, \mnras,
  477, 4856

\bibitem[{{Ahn} {et~al.}(2017){Ahn}, {Seth}, {den Brok}, {Strader},
  {Baumgardt}, {van den Bosch}, {Chilingarian}, {Frank}, {Hilker}, {McDermid},
  {Mieske}, {Romanowsky}, {Spitler}, {Brodie}, {Neumayer}, \&
  {Walsh}}]{Ahn2017}
{Ahn}, C.~P., {Seth}, A.~C., {den Brok}, M., {et~al.} 2017, \apj, 839, 72

\bibitem[{{Ahn} {et~al.}(2018){Ahn}, {Seth}, {Cappellari}, {Krajnovi{\'c}},
  {Strader}, {Voggel}, {Walsh}, {Bahramian}, {Baumgardt}, {Brodie},
  {Chilingarian}, {Chomiuk}, {den Brok}, {Frank}, {Hilker}, {McDermid},
  {Mieske}, {Neumayer}, {Nguyen}, {Pechetti}, {Romanowsky}, \&
  {Spitler}}]{Ahn2018}
{Ahn}, C.~P., {Seth}, A.~C., {Cappellari}, M., {et~al.} 2018, \apj, 858, 102

\bibitem[{{Antonini} {et~al.}(2015){Antonini}, {Barausse}, \&
  {Silk}}]{Antonini2015}
{Antonini}, F., {Barausse}, E., \& {Silk}, J. 2015, \apjl, 806, L8

\bibitem[{{Astropy Collaboration} {et~al.}(2013){Astropy Collaboration},
  {Robitaille}, {Tollerud}, {Greenfield}, {Droettboom}, {Bray}, {Aldcroft},
  {Davis}, {Ginsburg}, {Price-Whelan}, {Kerzendorf}, {Conley}, {Crighton},
  {Barbary}, {Muna}, {Ferguson}, {Grollier}, {Parikh}, {Nair}, {Unther},
  {Deil}, {Woillez}, {Conseil}, {Kramer}, {Turner}, {Singer}, {Fox}, {Weaver},
  {Zabalza}, {Edwards}, {Azalee Bostroem}, {Burke}, {Casey}, {Crawford},
  {Dencheva}, {Ely}, {Jenness}, {Labrie}, {Lim}, {Pierfederici}, {Pontzen},
  {Ptak}, {Refsdal}, {Servillat}, \& {Streicher}}]{astropy:2013}
{Astropy Collaboration}, {Robitaille}, T.~P., {Tollerud}, E.~J., {et~al.} 2013,
  aap, 558, A33

\bibitem[{{Barber} {et~al.}(2016){Barber}, {Schaye}, {Bower}, {Crain},
  {Schaller}, \& {Theuns}}]{Barber2016}
{Barber}, C., {Schaye}, J., {Bower}, R.~G., {et~al.} 2016, \mnras, 460, 1147

\bibitem[{{Baumgardt}(2017)}]{Baumgardt2017}
{Baumgardt}, H. 2017, \mnras, 464, 2174

\bibitem[{{Baumgardt} \& {Hilker}(2018)}]{Baumgardt2018}
{Baumgardt}, H., \& {Hilker}, M. 2018, \mnras, 478, 1520

\bibitem[{{Baumgardt} \& {Makino}(2003)}]{Baumgardt2003}
{Baumgardt}, H., \& {Makino}, J. 2003, \mnras, 340, 227

\bibitem[{{Bekki} {et~al.}(2001){Bekki}, {Couch}, \& {Drinkwater}}]{Bekki2001}
{Bekki}, K., {Couch}, W.~J., \& {Drinkwater}, M.~J. 2001, \apjl, 552, L105

\bibitem[{{Bekki} {et~al.}(2003){Bekki}, {Couch}, {Drinkwater}, \&
  {Shioya}}]{Bekki2003}
{Bekki}, K., {Couch}, W.~J., {Drinkwater}, M.~J., \& {Shioya}, Y. 2003, \mnras,
  344, 399

\bibitem[{{Blanton} {et~al.}(2005){Blanton}, {Lupton}, {Schlegel}, {Strauss},
  {Brinkmann}, {Fukugita}, \& {Loveday}}]{Blanton2005}
{Blanton}, M.~R., {Lupton}, R.~H., {Schlegel}, D.~J., {et~al.} 2005, \apj, 631,
  208

\bibitem[{{Brodie} {et~al.}(2011){Brodie}, {Romanowsky}, {Strader}, \&
  {Forbes}}]{Brodie2011}
{Brodie}, J.~P., {Romanowsky}, A.~J., {Strader}, J., \& {Forbes}, D.~A. 2011,
  \aj, 142, 199

\bibitem[{{Bruzual} \& {Charlot}(2003)}]{BC2003}
{Bruzual}, G., \& {Charlot}, S. 2003, \mnras, 344, 1000

\bibitem[{{Caldwell} {et~al.}(2011){Caldwell}, {Schiavon}, {Morrison}, {Rose},
  \& {Harding}}]{Caldwell2011}
{Caldwell}, N., {Schiavon}, R., {Morrison}, H., {Rose}, J.~A., \& {Harding}, P.
  2011, \aj, 141, 61

\bibitem[{{Chilingarian} {et~al.}(2018){Chilingarian}, {Katkov}, {Zolotukhin},
  {Grishin}, {Beletsky}, {Boutsia}, \& {Osip}}]{Chilingarian2018}
{Chilingarian}, I.~V., {Katkov}, I.~Y., {Zolotukhin}, I.~Y., {et~al.} 2018,
  \apj, 863, 1

\bibitem[{{Chilingarian} \& {Mamon}(2008)}]{Chili2008b}
{Chilingarian}, I.~V., \& {Mamon}, G.~A. 2008, \mnras, 385, L83

\bibitem[{{C{\^o}t{\'e}} {et~al.}(2006){C{\^o}t{\'e}}, {Piatek}, {Ferrarese},
  {Jord{\'a}n}, {Merritt}, {Peng}, {Ha{\c s}egan}, {Blakeslee}, {Mei}, {West},
  {Milosavljevi{\'c}}, \& {Tonry}}]{Cote2006}
{C{\^o}t{\'e}}, P., {Piatek}, S., {Ferrarese}, L., {et~al.} 2006, \apjs, 165,
  57

\bibitem[{{Da Rocha} {et~al.}(2011){Da Rocha}, {Mieske}, {Georgiev}, {Hilker},
  {Ziegler}, \& {Mendes de Oliveira}}]{DaRocha2011}
{Da Rocha}, C., {Mieske}, S., {Georgiev}, I.~Y., {et~al.} 2011, \aap, 525, A86

\bibitem[{{Dabringhausen} {et~al.}(2009){Dabringhausen}, {Kroupa}, \&
  {Baumgardt}}]{Dabringhausen2009}
{Dabringhausen}, J., {Kroupa}, P., \& {Baumgardt}, H. 2009, \mnras, 394, 1529

\bibitem[{{Dabringhausen} {et~al.}(2012){Dabringhausen}, {Kroupa},
  {Pflamm-Altenburg}, \& {Mieske}}]{Dabringhausen2012}
{Dabringhausen}, J., {Kroupa}, P., {Pflamm-Altenburg}, J., \& {Mieske}, S.
  2012, \apj, 747, 72

\bibitem[{{De Lucia} \& {Blaizot}(2007)}]{DeLucia2007}
{De Lucia}, G., \& {Blaizot}, J. 2007, \mnras, 375, 2

\bibitem[{{den Brok} {et~al.}(2015){den Brok}, {Seth}, {Barth}, {Carson},
  {Neumayer}, {Cappellari}, {Debattista}, {Ho}, {Hood}, \&
  {McDermid}}]{denBrok2015}
{den Brok}, M., {Seth}, A.~C., {Barth}, A.~J., {et~al.} 2015, \apj, 809, 101

\bibitem[{{Drinkwater} {et~al.}(2003){Drinkwater}, {Gregg}, {Hilker}, {Bekki},
  {Couch}, {Ferguson}, {Jones}, \& {Phillipps}}]{Drinkwater2003}
{Drinkwater}, M.~J., {Gregg}, M.~D., {Hilker}, M., {et~al.} 2003, \nat, 423,
  519

\bibitem[{{Drinkwater} {et~al.}(2000){Drinkwater}, {Jones}, {Gregg}, \&
  {Phillipps}}]{Drinkwater2000}
{Drinkwater}, M.~J., {Jones}, J.~B., {Gregg}, M.~D., \& {Phillipps}, S. 2000,
  \pasa, 17, 227

\bibitem[{{Ferrarese} {et~al.}(2006){Ferrarese}, {C{\^o}t{\'e}}, {Jord{\'a}n},
  {Peng}, {Blakeslee}, {Piatek}, {Mei}, {Merritt}, {Milosavljevi{\'c}},
  {Tonry}, \& {West}}]{Ferrarese2006}
{Ferrarese}, L., {C{\^o}t{\'e}}, P., {Jord{\'a}n}, A., {et~al.} 2006, \apjs,
  164, 334

\bibitem[{{Ferrarese} {et~al.}(2016){Ferrarese}, {C{\^o}t{\'e}},
  {S{\'a}nchez-Janssen}, {Roediger}, {McConnachie}, {Durrell}, {MacArthur},
  {Blakeslee}, {Duc}, {Boissier}, {Boselli}, {Courteau}, {Cuillandre},
  {Emsellem}, {Gwyn}, {Guhathakurta}, {Jord{\'a}n}, {Lan{\c c}on}, {Liu},
  {Mei}, {Mihos}, {Navarro}, {Peng}, {Puzia}, {Taylor}, {Toloba}, \&
  {Zhang}}]{Ferrarese2016}
{Ferrarese}, L., {C{\^o}t{\'e}}, P., {S{\'a}nchez-Janssen}, R., {et~al.} 2016,
  \apj, 824, 10

\bibitem[{{Firth} {et~al.}(2009){Firth}, {Evstigneeva}, \&
  {Drinkwater}}]{Firth2009}
{Firth}, P., {Evstigneeva}, E.~A., \& {Drinkwater}, M.~J. 2009, \mnras, 394,
  1801

\bibitem[{{Forbes} {et~al.}(2014){Forbes}, {Norris}, {Strader}, {Romanowsky},
  {Pota}, {Kannappan}, {Brodie}, \& {Huxor}}]{Forbes2014}
{Forbes}, D.~A., {Norris}, M.~A., {Strader}, J., {et~al.} 2014, \mnras, 444,
  2993

\bibitem[{{French} {et~al.}(2016){French}, {Arcavi}, \&
  {Zabludoff}}]{French2016}
{French}, K.~D., {Arcavi}, I., \& {Zabludoff}, A. 2016, \apjl, 818, L21

\bibitem[{{Gond{\'a}n} {et~al.}(2018{\natexlab{a}}){Gond{\'a}n}, {Kocsis},
  {Raffai}, \& {Frei}}]{Gondan2018b}
{Gond{\'a}n}, L., {Kocsis}, B., {Raffai}, P., \& {Frei}, Z. 2018{\natexlab{a}},
  \apj, 855, 34

\bibitem[{{Gond{\'a}n} {et~al.}(2018{\natexlab{b}}){Gond{\'a}n}, {Kocsis},
  {Raffai}, \& {Frei}}]{Gondan2018a}
---. 2018{\natexlab{b}}, \apj, 860, 5

\bibitem[{{Graham} \& {Spitler}(2009)}]{Graham2009}
{Graham}, A.~W., \& {Spitler}, L.~R. 2009, \mnras, 397, 2148

\bibitem[{{Haghi} {et~al.}(2017){Haghi}, {Khalaj}, {Hasani Zonoozi}, \&
  {Kroupa}}]{Haghi2017}
{Haghi}, H., {Khalaj}, P., {Hasani Zonoozi}, A., \& {Kroupa}, P. 2017, \apj,
  839, 60

\bibitem[{{Hilker}(2006)}]{Hilker2006}
{Hilker}, M. 2006, ArXiv Astrophysics e-prints, astro-ph/0605447

\bibitem[{{Hilker}(2009)}]{Hilker2009}
{Hilker}, M. 2009, in Reviews in Modern Astronomy, Vol.~21, Reviews in Modern
  Astronomy, ed. S.~{R{\"o}ser}, 199--213

\bibitem[{{Hilker} {et~al.}(2007){Hilker}, {Baumgardt}, {Infante},
  {Drinkwater}, {Evstigneeva}, \& {Gregg}}]{Hilker2007}
{Hilker}, M., {Baumgardt}, H., {Infante}, L., {et~al.} 2007, \aap, 463, 119

\bibitem[{{Hilker} {et~al.}(1999){Hilker}, {Infante}, {Vieira},
  {Kissler-Patig}, \& {Richtler}}]{Hilker1999}
{Hilker}, M., {Infante}, L., {Vieira}, G., {Kissler-Patig}, M., \& {Richtler},
  T. 1999, \aaps, 134, 75

\bibitem[{{Janz} {et~al.}(2015){Janz}, {Forbes}, {Norris}, {Strader}, {Penny},
  {Fagioli}, \& {Romanowsky}}]{Janz2015}
{Janz}, J., {Forbes}, D.~A., {Norris}, M.~A., {et~al.} 2015, \mnras, 449, 1716

\bibitem[{{Janz} {et~al.}(2016){Janz}, {Norris}, {Forbes}, {Huxor},
  {Romanowsky}, {Frank}, {Escudero}, {Faifer}, {Forte}, {Kannappan},
  {Maraston}, {Brodie}, {Strader}, \& {Thompson}}]{Janz2016}
{Janz}, J., {Norris}, M.~A., {Forbes}, D.~A., {et~al.} 2016, \mnras, 456, 617

\bibitem[{{Kelly}(2007)}]{Kelly2007}
{Kelly}, B.~C. 2007, \apj, 665, 1489

\bibitem[{{Kimmig} {et~al.}(2015){Kimmig}, {Seth}, {Ivans}, {Strader},
  {Caldwell}, {Anderton}, \& {Gregersen}}]{Kimmig2015}
{Kimmig}, B., {Seth}, A., {Ivans}, I.~I., {et~al.} 2015, \aj, 149, 53

\bibitem[{{Kocsis} \& {Levin}(2012)}]{Kocsis2012}
{Kocsis}, B., \& {Levin}, J. 2012, \prd, 85, 123005

\bibitem[{{Kroupa}(2002)}]{Kroupa2002}
{Kroupa}, P. 2002, Science, 295, 82

\bibitem[{{Kruijssen} \& {Mieske}(2009)}]{Krui2009}
{Kruijssen}, J.~M.~D., \& {Mieske}, S. 2009, \aap, 500, 785

\bibitem[{{Law-Smith} {et~al.}(2017){Law-Smith}, {Ramirez-Ruiz}, {Ellison}, \&
  {Foley}}]{LawSmith2017}
{Law-Smith}, J., {Ramirez-Ruiz}, E., {Ellison}, S.~L., \& {Foley}, R.~J. 2017,
  \apj, 850, 22

\bibitem[{{Maraston}(2005)}]{Maraston2005}
{Maraston}, C. 2005, \mnras, 362, 799

\bibitem[{{Mieske} {et~al.}(2013){Mieske}, {Frank}, {Baumgardt},
  {L{\"u}tzgendorf}, {Neumayer}, \& {Hilker}}]{Mieske2013}
{Mieske}, S., {Frank}, M.~J., {Baumgardt}, H., {et~al.} 2013, \aap, 558, A14

\bibitem[{{Mieske} {et~al.}(2004){Mieske}, {Hilker}, \& {Infante}}]{Mieske2004}
{Mieske}, S., {Hilker}, M., \& {Infante}, L. 2004, \aap, 418, 445

\bibitem[{{Mieske} {et~al.}(2012){Mieske}, {Hilker}, \& {Misgeld}}]{Mieske2012}
{Mieske}, S., {Hilker}, M., \& {Misgeld}, I. 2012, \aap, 537, A3

\bibitem[{{Mieske} {et~al.}(2008){Mieske}, {Hilker}, {Jord{\'a}n}, {Infante},
  {Kissler-Patig}, {Rejkuba}, {Richtler}, {C{\^o}t{\'e}}, {Baumgardt}, {West},
  {Ferrarese}, \& {Peng}}]{Mieske2008}
{Mieske}, S., {Hilker}, M., {Jord{\'a}n}, A., {et~al.} 2008, \aap, 487, 921

\bibitem[{{Minniti} {et~al.}(1998){Minniti}, {Kissler-Patig}, {Goudfrooij}, \&
  {Meylan}}]{Minniti1998}
{Minniti}, D., {Kissler-Patig}, M., {Goudfrooij}, P., \& {Meylan}, G. 1998,
  \aj, 115, 121

\bibitem[{{Mockler} {et~al.}(2018){Mockler}, {Guillochon}, \&
  {Ramirez-Ruiz}}]{Mockler2018}
{Mockler}, B., {Guillochon}, J., \& {Ramirez-Ruiz}, E. 2018, ArXiv e-prints,
  arXiv:1801.08221

\bibitem[{{Mu{\~n}oz} {et~al.}(2015){Mu{\~n}oz}, {Eigenthaler}, {Puzia},
  {Taylor}, {Ordenes-Brice{\~n}o}, {Alamo-Mart{\'{\i}}nez}, {Ribbeck},
  {{\'A}ngel}, {Capaccioli}, {C{\^o}t{\'e}}, {Ferrarese}, {Galaz}, {Hempel},
  {Hilker}, {Jord{\'a}n}, {Lan{\c c}on}, {Mieske}, {Paolillo}, {Richtler},
  {S{\'a}nchez-Janssen}, \& {Zhang}}]{Munoz2015}
{Mu{\~n}oz}, R.~P., {Eigenthaler}, P., {Puzia}, T.~H., {et~al.} 2015, \apjl,
  813, L15

\bibitem[{{Murray}(2009)}]{Murray2009}
{Murray}, N. 2009, \apj, 691, 946

\bibitem[{{Nguyen} {et~al.}(2018){Nguyen}, {Seth}, {Neumayer}, {Kamann},
  {Voggel}, {Cappellari}, {Picotti}, {Nguyen}, {B{\"o}ker}, {Debattista},
  {Caldwell}, {McDermid}, {Bastian}, {Ahn}, \& {Pechetti}}]{Nguyen2018}
{Nguyen}, D.~D., {Seth}, A.~C., {Neumayer}, N., {et~al.} 2018, \apj, 858, 118

\bibitem[{{Norris} {et~al.}(2015){Norris}, {Escudero}, {Faifer}, {Kannappan},
  {Forte}, \& {van den Bosch}}]{Norris2015}
{Norris}, M.~A., {Escudero}, C.~G., {Faifer}, F.~R., {et~al.} 2015, \mnras,
  451, 3615

\bibitem[{{Norris} \& {Kannappan}(2011)}]{Norris2011}
{Norris}, M.~A., \& {Kannappan}, S.~J. 2011, \mnras, 414, 739

\bibitem[{{Norris} {et~al.}(2014){Norris}, {Kannappan}, {Forbes}, {Romanowsky},
  {Brodie}, {Faifer}, {Huxor}, {Maraston}, {Moffett}, {Penny}, {Pota},
  {Smith-Castelli}, {Strader}, {Bradley}, {Eckert}, {Fohring}, {McBride},
  {Stark}, \& {Vaduvescu}}]{Norris2014}
{Norris}, M.~A., {Kannappan}, S.~J., {Forbes}, D.~A., {et~al.} 2014, \mnras,
  443, 1151

\bibitem[{{O'Leary} {et~al.}(2009){O'Leary}, {Kocsis}, \& {Loeb}}]{OLeary2009}
{O'Leary}, R.~M., {Kocsis}, B., \& {Loeb}, A. 2009, \mnras, 395, 2127

\bibitem[{{Ordenes-Brice{\~n}o} {et~al.}(2018){Ordenes-Brice{\~n}o}, {Puzia},
  {Eigenthaler}, {Taylor}, {Mu{\~n}oz}, {Zhang}, {Alamo-Mart{\'{\i}}nez},
  {Ribbeck}, {Grebel}, {{\'A}ngel}, {C{\^o}t{\'e}}, {Ferrarese}, {Hilker},
  {Lan{\c c}on}, {Mieske}, {Miller}, {Rong}, \&
  {S{\'a}nchez-Janssen}}]{Ordenes2018}
{Ordenes-Brice{\~n}o}, Y., {Puzia}, T.~H., {Eigenthaler}, P., {et~al.} 2018,
  \apj, 860, 4

\bibitem[{Pedregosa {et~al.}(2011)Pedregosa, Varoquaux, Gramfort, Michel,
  Thirion, Grisel, Blondel, Prettenhofer, Weiss, Dubourg, Vanderplas, Passos,
  Cournapeau, Brucher, Perrot, \& Duchesnay}]{scikit-learn}
Pedregosa, F., Varoquaux, G., Gramfort, A., {et~al.} 2011, Journal of Machine
  Learning Research, 12, 2825

\bibitem[{{Pfeffer} \& {Baumgardt}(2013)}]{Pfeffer2013}
{Pfeffer}, J., \& {Baumgardt}, H. 2013, \mnras, 433, 1997

\bibitem[{{Pfeffer} {et~al.}(2014){Pfeffer}, {Griffen}, {Baumgardt}, \&
  {Hilker}}]{Pfeffer2014}
{Pfeffer}, J., {Griffen}, B.~F., {Baumgardt}, H., \& {Hilker}, M. 2014, \mnras,
  444, 3670

\bibitem[{{Pfeffer} {et~al.}(2016){Pfeffer}, {Hilker}, {Baumgardt}, \&
  {Griffen}}]{Pfeffer2016}
{Pfeffer}, J., {Hilker}, M., {Baumgardt}, H., \& {Griffen}, B.~F. 2016, \mnras,
  458, 2492

\bibitem[{{Price-Whelan} {et~al.}(2018){Price-Whelan}, {Sip{'{o}}cz},
  {G{"u}nther}, {Lim}, {Crawford}, {Conseil}, {Shupe}, {Craig}, {Dencheva},
  {Ginsburg}, {VanderPlas}, {Bradley}, {P{'e}rez-Su{'a}rez}, {de Val-Borro},
  {Paper Contributors}, {Aldcroft}, {Cruz}, {Robitaille}, {Tollerud},
  {Coordination Committee}, {Ardelean}, {Babej}, {Bach}, {Bachetti}, {Bakanov},
  {Bamford}, {Barentsen}, {Barmby}, {Baumbach}, {Berry}, {Biscani}, {Boquien},
  {Bostroem}, {Bouma}, {Brammer}, {Bray}, {Breytenbach}, {Buddelmeijer},
  {Burke}, {Calderone}, {Cano Rodr{'i}guez}, {Cara}, {Cardoso}, {Cheedella},
  {Copin}, {Corrales}, {Crichton}, {D{ extquoteright}Avella}, {Deil},
  {Depagne}, {Dietrich}, {Donath}, {Droettboom}, {Earl}, {Erben}, {Fabbro},
  {Ferreira}, {Finethy}, {Fox}, {Garrison}, {Gibbons}, {Goldstein}, {Gommers},
  {Greco}, {Greenfield}, {Groener}, {Grollier}, {Hagen}, {Hirst}, {Homeier},
  {Horton}, {Hosseinzadeh}, {Hu}, {Hunkeler}, {Ivezi{'c}}, {Jain}, {Jenness},
  {Kanarek}, {Kendrew}, {Kern}, {Kerzendorf}, {Khvalko}, {King}, {Kirkby},
  {Kulkarni}, {Kumar}, {Lee}, {Lenz}, {Littlefair}, {Ma}, {Macleod},
  {Mastropietro}, {McCully}, {Montagnac}, {Morris}, {Mueller}, {Mumford},
  {Muna}, {Murphy}, {Nelson}, {Nguyen}, {Ninan}, {N{"o}the}, {Ogaz}, {Oh},
  {Parejko}, {Parley}, {Pascual}, {Patil}, {Patil}, {Plunkett}, {Prochaska},
  {Rastogi}, {Reddy Janga}, {Sabater}, {Sakurikar}, {Seifert}, {Sherbert},
  {Sherwood-Taylor}, {Shih}, {Sick}, {Silbiger}, {Singanamalla}, {Singer},
  {Sladen}, {Sooley}, {Sornarajah}, {Streicher}, {Teuben}, {Thomas},
  {Tremblay}, {Turner}, {Terr{'o}n}, {van Kerkwijk}, {de la Vega}, {Watkins},
  {Weaver}, {Whitmore}, {Woillez}, {Zabalza}, \& {Contributors}}]{astropy:2018}
{Price-Whelan}, A.~M., {Sip{'{o}}cz}, B.~M., {G{"u}nther}, H.~M., {et~al.}
  2018, aj, 156, 123

\bibitem[{{Rasskazov} \& {Kocsis}(2018)}]{Rasskazov2018}
{Rasskazov}, A., \& {Kocsis}, B. 2018, in preparation

\bibitem[{{Reines} {et~al.}(2013){Reines}, {Greene}, \& {Geha}}]{Reines2013}
{Reines}, A.~E., {Greene}, J.~E., \& {Geha}, M. 2013, \apj, 775, 116

\bibitem[{{Rejkuba} {et~al.}(2007){Rejkuba}, {Dubath}, {Minniti}, \&
  {Meylan}}]{Rejkuba2007}
{Rejkuba}, M., {Dubath}, P., {Minniti}, D., \& {Meylan}, G. 2007, \aap, 469,
  147

\bibitem[{{Saglia} {et~al.}(2016){Saglia}, {Opitsch}, {Erwin}, {Thomas},
  {Beifiori}, {Fabricius}, {Mazzalay}, {Nowak}, {Rusli}, \&
  {Bender}}]{Saglia2016}
{Saglia}, R.~P., {Opitsch}, M., {Erwin}, P., {et~al.} 2016, \apj, 818, 47

\bibitem[{{Sandoval} {et~al.}(2015){Sandoval}, {Vo}, {Romanowsky}, {Strader},
  {Choi}, {Jennings}, {Conroy}, {Brodie}, {Foster}, {Villaume}, {Norris},
  {Janz}, \& {Forbes}}]{Sandoval2015}
{Sandoval}, M.~A., {Vo}, R.~P., {Romanowsky}, A.~J., {et~al.} 2015, \apjl, 808,
  L32

\bibitem[{{Schweizer} {et~al.}(2018){Schweizer}, {Seitzer}, {Whitmore},
  {Kelson}, \& {Villanueva}}]{Schweizer2018}
{Schweizer}, F., {Seitzer}, P., {Whitmore}, B.~C., {Kelson}, D.~D., \&
  {Villanueva}, E.~V. 2018, \apj, 853, 54

\bibitem[{{Seth} {et~al.}(2006){Seth}, {Dalcanton}, {Hodge}, \&
  {Debattista}}]{Seth2006}
{Seth}, A.~C., {Dalcanton}, J.~J., {Hodge}, P.~W., \& {Debattista}, V.~P. 2006,
  \aj, 132, 2539

\bibitem[{{Seth} {et~al.}(2014){Seth}, {van den Bosch}, {Mieske}, {Baumgardt},
  {Brok}, {Strader}, {Neumayer}, {Chilingarian}, {Hilker}, {McDermid},
  {Spitler}, {Brodie}, {Frank}, \& {Walsh}}]{Seth2014}
{Seth}, A.~C., {van den Bosch}, R., {Mieske}, S., {et~al.} 2014, \nat, 513, 398

\bibitem[{Shankar {et~al.}(2004)Shankar, Salucci, Granato, De~Zotti, \&
  Danese}]{Shankar2004}
Shankar, F., Salucci, P., Granato, G.~L., De~Zotti, G., \& Danese, L. 2004,
  \mnras, 354, 1020

\bibitem[{{Sollima} \& {Baumgardt}(2017)}]{Sollima2017}
{Sollima}, A., \& {Baumgardt}, H. 2017, \mnras, 471, 3668

\bibitem[{{Steinmetz} \& {Navarro}(2002)}]{Steinmetz2002}
{Steinmetz}, M., \& {Navarro}, J.~F. 2002, \na, 7, 155

\bibitem[{{Strader} {et~al.}(2011){Strader}, {Caldwell}, \&
  {Seth}}]{Strader2011}
{Strader}, J., {Caldwell}, N., \& {Seth}, A.~C. 2011, \aj, 142, 8

\bibitem[{{Strader} {et~al.}(2013){Strader}, {Seth}, {Forbes}, {Fabbiano},
  {Romanowsky}, {Brodie}, {Conroy}, {Caldwell}, {Pota}, {Usher}, \&
  {Arnold}}]{Strader2013}
{Strader}, J., {Seth}, A.~C., {Forbes}, D.~A., {et~al.} 2013, \apjl, 775, L6

\bibitem[{{Taylor} {et~al.}(2015){Taylor}, {Puzia}, {Gomez}, \&
  {Woodley}}]{Taylor2015}
{Taylor}, M.~A., {Puzia}, T.~H., {Gomez}, M., \& {Woodley}, K.~A. 2015, \apj,
  805, 65

\bibitem[{{Taylor} {et~al.}(2010){Taylor}, {Puzia}, {Harris}, {Harris},
  {Kissler-Patig}, \& {Hilker}}]{Taylor2010}
{Taylor}, M.~A., {Puzia}, T.~H., {Harris}, G.~L., {et~al.} 2010, \apj, 712,
  1191

\bibitem[{{Tremmel} {et~al.}(2018){Tremmel}, {Governato}, {Volonteri},
  {Pontzen}, \& {Quinn}}]{Tremmel2018}
{Tremmel}, M., {Governato}, F., {Volonteri}, M., {Pontzen}, A., \& {Quinn},
  T.~R. 2018, \apj, 857, L22

\bibitem[{Tsang(2013)}]{Tsang2013}
Tsang, D. 2013, The Astrophysical Journal, 777, 103

\bibitem[{{Turner} {et~al.}(2012){Turner}, {C{\^o}t{\'e}}, {Ferrarese},
  {Jord{\'a}n}, {Blakeslee}, {Mei}, {Peng}, \& {West}}]{Turner2012}
{Turner}, M.~L., {C{\^o}t{\'e}}, P., {Ferrarese}, L., {et~al.} 2012, \apjs,
  203, 5

\bibitem[{Van Der~Walt {et~al.}(2011)Van Der~Walt, Colbert, \&
  Varoquaux}]{van2011numpy}
Van Der~Walt, S., Colbert, S.~C., \& Varoquaux, G. 2011, Computing in Science
  \& Engineering, 13, 22

\bibitem[{{Vesperini} \& {Heggie}(1997)}]{Vesperini1997}
{Vesperini}, E., \& {Heggie}, D.~C. 1997, \mnras, 289, 898

\bibitem[{{Villaume} {et~al.}(2017){Villaume}, {Brodie}, {Conroy},
  {Romanowsky}, \& {van Dokkum}}]{Villaume2017}
{Villaume}, A., {Brodie}, J., {Conroy}, C., {Romanowsky}, A.~J., \& {van
  Dokkum}, P. 2017, \apjl, 850, L14

\bibitem[{{Voggel} {et~al.}(2018){Voggel}, {Seth}, {Neumayer}, {Mieske},
  {Chilingarian}, {Ahn}, {Baumgardt}, {Hilker}, {Nguyen}, {Romanowsky},
  {Walsh}, {den Brok}, \& {Strader}}]{Voggel2018}
{Voggel}, K.~T., {Seth}, A.~C., {Neumayer}, N., {et~al.} 2018, \apj, 858, 20

\bibitem[{{Volonteri} {et~al.}(2016){Volonteri}, {Dubois}, {Pichon}, \&
  {Devriendt}}]{Volonteri2016}
{Volonteri}, M., {Dubois}, Y., {Pichon}, C., \& {Devriendt}, J. 2016, \mnras,
  460, 2979

\bibitem[{{Volonteri} {et~al.}(2008){Volonteri}, {Haardt}, \&
  {G{\"u}ltekin}}]{Volonteri2008}
{Volonteri}, M., {Haardt}, F., \& {G{\"u}ltekin}, K. 2008, \mnras, 384, 1387

\bibitem[{{Walcher} {et~al.}(2006){Walcher}, {B{\"o}ker}, {Charlot}, {Ho},
  {Rix}, {Rossa}, {Shields}, \& {van der Marel}}]{Walcher2006}
{Walcher}, C.~J., {B{\"o}ker}, T., {Charlot}, S., {et~al.} 2006, \apj, 649, 692

\bibitem[{{Wevers} {et~al.}(2017){Wevers}, {van Velzen}, {Jonker}, {Stone},
  {Hung}, {Onori}, {Gezari}, \& {Blagorodnova}}]{Wevers2017}
{Wevers}, T., {van Velzen}, S., {Jonker}, P.~G., {et~al.} 2017, \mnras, 471,
  1694

\bibitem[{{Zhang} {et~al.}(2015){Zhang}, {Peng}, {C{\^o}t{\'e}}, {Liu},
  {Ferrarese}, {Cuillandre}, {Caldwell}, {Gwyn}, {Jord{\'a}n}, {Lan{\c c}on},
  {Li}, {Mu{\~n}oz}, {Puzia}, {Bekki}, {Blakeslee}, {Boselli}, {Drinkwater},
  {Duc}, {Durrell}, {Emsellem}, {Firth}, \& {S{\'a}nchez-Janssen}}]{Zhang2015}
{Zhang}, H.-X., {Peng}, E.~W., {C{\^o}t{\'e}}, P., {et~al.} 2015, \apj, 802, 30

\end{thebibliography}




\end{document}